 \documentclass[journal]{IEEEtran}
\newif\ifCLASSOPTIONromanappendices \CLASSOPTIONromanappendicestrue
\usepackage{amsmath,amsfonts,graphicx,booktabs}
 \usepackage{url}
  \usepackage{color}

\newcommand{\Trace}{\operatorname{Tr}}
\newtheorem{lemma}{Lemma}
\usepackage{cite}
\usepackage{algpseudocode}
\usepackage{algorithm}
\usepackage{epstopdf}
\usepackage{hyperref}
\usepackage{fix2col}

  \newlength{\figwidth}
   \newcommand{\revtsix}[1]{{\color{blue}#1}}
   \renewcommand{\revtsix}[1]{#1}
     \newcommand{\revtsev}[1]{{\color{blue}#1}}
      \renewcommand{\revtsev}[1]{#1}
     \newcommand{\revteight}[1]{{\color{magenta}#1}}
      \renewcommand{\revteight}[1]{#1}
      \newcommand{\revtnine}[1]{{\color{blue}#1}}
            \renewcommand{\revtnine}[1]{#1}
            \newcommand{\revtten}[1]{{\color{red}#1}}
             \renewcommand{\revtten}[1]{#1}
   \newcommand{\complexunit}{\mathsf{i}}
   
   \newcommand{\changet}[1]{{\color{magenta}#1}}
   \newcommand{\changett}[1]{{\color{magenta}#1}}
    \renewcommand{\changet}[1]{#1}
        \renewcommand{\changett}[1]{#1}

   \newcommand{\newchanget}[1]{{\color{blue}#1}}
   \newcommand{\newchangem}[1]{{\color{red}#1}}
 \renewcommand{\newchangem}[1]{{\color{blue}#1}}
 \renewcommand{\newchanget}[1]{#1}
  \renewcommand{\newchangem}[1]{#1}

\title{Coordinate Update Algorithms for Robust Power Loading for the %TDD 
\newchanget{MU-}MISO Downlink with Outage Constraints
\thanks{Manuscript accepted and to appear in IEEE Transactions on Signal Processing, 2016. This work was supported in part by  the
Natural Sciences and Engineering Research Council of Canada under grant RGPIN-2015-06631. 
A preliminary 
\changet{exposition of some of the ideas in this work, in the context of a slightly different
receiver structure,} appears in 
\textit{Proc.\ 2013 IEEE  Int.\ Conf.\ Acoust.\ Speech, Signal Process.}}}

\author{Foad Sohrabi and Timothy N. Davidson\textsuperscript{*}
\thanks{F. Sohrabi was with, and T. N. Davidson is with, the Department of Electrical and Computer Engineering, McMaster University, 
1280 Main Street West, Hamilton, Ontario, Canada, L8S 4K1. 
Tel: +1 905 525 9140, Ext.~27352. Fax: +1 905 521 2922. Email:  davidson@mcmaster.ca. F. Sohrabi is now with the Department of Electrical and Computer Engineering at the
University of Toronto. Email: foad.sohrabi@mail.utoronto.ca.}}

\begin{document}
%\ninept 
\sloppy
\maketitle   

\begin{abstract}
%this {\color{red} will need to be refined. do we drop tdd from title, and just say that it is 
%appropriate for TDD? changing channels/ delayed feedback remains a problem. 
%perhaps we need to argue for a tracker and the Gaussian error in the tracker.}
%
We consider the problem of power allocation for the single-cell 
\newchanget{multi-user (MU)} multiple-input single-output (MISO)
downlink with quality-of-service (QoS) constraints. The base station acquires an estimate of the channel\newchanget{s}
and, for a given beamforming structure,
designs the power allocation so as to minimize the total transmission power required to ensure that target
signal-to-interference-and-noise ratios at the receivers are met, subject to a specified outage probability.
We consider scenarios in which the  errors in the base station's channel estimates 
can be modelled as being zero-mean and Gaussian.  Such a model is particularly suitable for 
time division duplex (TDD) systems with quasi-static channels, in which the base station estimates the 
channel during the uplink phase. 
Under that model, we %obtain 
\revtsix{employ} a precise deterministic characterization
of the outage probability
%, and mildly conservative approximations thereof.
\revtsev{to transform the chance-constrained formulation to a deterministic one.
Although that deterministic formulation is not convex, we develop a coordinate
descent algorithm that
%, through connections with the framework of interference functions,
can be shown to converge to a globally optimal solution when the starting point is feasible.
Insight into the structure of the deterministic formulation yields approximations that
result in coordinate update algorithms with good performance and significantly lower
computational cost. 
%As illustrated by simulation results, the 
The proposed
algorithms provide better performance than existing robust power loading algorithms 
that are based on tractable conservative approximations, and can even provide
better performance than robust precoding algorithms based on such approximations.}
%
%
%Although the resulting deterministic formulations of the design problem are not convex, 
%\revtsix{connections with the framework of interference functions suggest the 
%development of a variety of coordinate update algorithms.
%The solutions thus obtained}
%%we have been able to obtain good solutions using straightforward coordinate update algorithms. In fact, these solutions 
%provide significantly better performance
%than the existing approaches, which are based on convex restrictions, because the proposed approximations are less conservative. By developing some approximations of the precise deterministic characterization of the outage probability, we develop algorithms that have good performance and much lower computational cost.
%
%
% in a time division duplex (TDD) system. In such systems, the base station (BS) acquires information about the channel state during the training component of the uplink phase. The resulting estimation errors are modelled probabilistically, and the receivers specify quality-of-service (QoS)
%constraints in terms of a target signal-to-interference-and-noise ratio  that is to be achieved with a given
%outage probability. For a fixed beamforming structure, we seek a power allocation that minimizes the transmission power required to satisfy the users' QoS requests. 
%
%The proposed approach to that problem begins with the observation that for TDD systems the channel estimation error at the base station can be modelled as being additive and Gaussian. 
%%
\end{abstract}
\begin{IEEEkeywords}
Broadcast channel, \revtsix{downlink beamforming, robust power loading,
chance constraints, interference functions.}
\end{IEEEkeywords}

%\renewcommand{\IEEEkeywordsname}{EDICS}
%\begin{IEEEkeywords}
%SPC--MISG, 
%SPC--INTF
%\end{IEEEkeywords}

%======================================================================
%===============Introduction Section =========================================
%======================================================================

 %\newpage
\section{Introduction}
%{\color{red} Intro; there is much to be smoothed out here.}

It has long been recognized that the  provision of multiple antennas at the transmitter of a
downlink system has the potential to significantly improve  the efficiency
with which messages can be communicated from the base station to the receivers; e.g., \cite{Winters_Salz_Gitlin,Weingarten_2006,gesbert2007shifting}. 
In scenarios in which the base station has perfect knowledge of the state of the channels to
each of the single-antenna receivers \newchanget{(and has independent messages to send to them)}, the dirty paper coding scheme is optimal in the sense it enables the system to achieve any rate tuple in the capacity region \cite{caire2003achievable,Weingarten_2006}. Since that scheme is 
%based on the principle of pre-subtraction of the interference that the base station knows it will induce at the receivers, and is 
quite complicated to implement,  a variety of simpler sequential interference pre-subtraction schemes, such as those based on Tomlinson-Harashima precoders \cite{windpassinger2004precoding,fung2007precoding,liu2008novel} and vector perturbation precoding \cite{hochwald2005vector} have been considered.
Even   simpler schemes based on linear precoding \cite{gesbert2007shifting,Rashid-Farrokhi_1998_COM,Beamforming_Bengtsson_2001,spencer2004zero}, have also been considered, and we will consider the linear case herein.

In the case of  fixed-rate traffic, 
\newchanget{one approach to the design of} the linear precoder is 
%typically designed  so as 
to 
%one way in which that potential can be realized is to
  %design a linear transmitter so as to
   minimize the power that is required to enable reliable
communication to each receiver at their specified target rate.
%; e.g., \cite{Rashid-Farrokhi_1998_COM,Beamforming_Bengtsson_2001}.
 For narrowband systems in which
the receivers have a single antenna, that quality-of-service (QoS) problem is equivalent to
minimizing the transmission power required to satisfy a signal-to-interference-and-noise (SINR)
constraint at each receiver; i.e., $\displaystyle{\min \mathsf{power}} \displaystyle{\text{ subject to }  \mathsf{SINR}_k \geq  {\gamma}_k}$; e.g., \cite{Rashid-Farrokhi_1998_COM,Beamforming_Bengtsson_2001}.
%Although linear transmitters are not optimal for that problem~\cite{Weingarten_2006}, they
%are relatively simple to implement. 
Under the assumption that the transmitter can be provided
with accurate channel state information (CSI), without expending a significant fraction of the channel resources,
optimal linear precoders for a variety of such quality-of-service (QoS) problems have been obtained;
e.g., \newchanget{\cite{Rashid-Farrokhi_1998_COM,Beamforming_Bengtsson_2001,Schubert_Boche_2004,Wiesel_2006_Fixed,Schubert_Boche_2007,Hunger_Joham_2010,Huang_Palomar_separable}}.

In practice, however, the CSI that can be made available at the
transmitter is imperfect, due to estimation errors, quantization, feedback delay, feedback errors, and other 
effects; e.g.,\cite{Caire_Jindal_Kobayashi_Ravindran,Xu_Andrews_Jafar}. 
%{\color{red} Xu Jafar Andrews} 
For the QoS problems that we will consider herein, 
a straightforward approach to dealing with the
resulting uncertainty in the CSI is  to perform the design as if the channel estimates were correct, but
to increase the SINR targets, $\gamma_k$, 
%beyond the level required for reliable communication with
%the chosen signalling scheme, 
in order to increase the likelihood that this ``mismatched'' design meets
the original requirements.  
A more sophisticated approach is to 
 incorporate a model for the uncertainty into the transmitter design.
%A straightforward approach to 
One approach to doing that is to adopt a bounded model for the uncertainty and to design a transmitter  
that satisfies the QoS requirements even for the worst case of the uncertainties admitted by the model; i.e., $\displaystyle{\min \mathsf{power}\text{ subject to } \mathsf{SINR}_k \geq  {\gamma}_k}$ for all admitted uncertainties; e.g., \cite{Michael_JSTSP,KKWong_robust_downlink_SDR,Vucic_Boche_2009,Michael_QoS_MSE}. 
In this paper we  consider an alternative approach in which the uncertainty is modelled probabilistically
and the
QoS requirements are to be satisfied up to a given probability of outage; i.e, $\displaystyle{\min \mathsf{power}  \text{ subject to } \operatorname{Pr}(\mathsf{SINR}_k \geq  {\gamma}_k) \geq 1-\epsilon_k}$;
e.g., \newchanget{\cite{K_Boyd_2002,Chalise_2007_CC_DL,Payaro_2007,Michael_Asilomar08,Vucic_Boche,Ken_outage}}.
%{\color{red}Check Ken Ma's paper for publication}

The focus of this paper will be on scenarios in which the uncertainty can be modelled as a zero-mean Gaussian 
random variable with a given covariance. These scenarios include the case of (single-cell) time-division duplex (TDD) systems operating in quasi-static channels, in which the dominant component of the uncertainty arises
from the channel estimation error \newchangem{of the (unbiased) estimator}
on the uplink, and systems in which the channel variation is tracked by the
transmitter using variants of the Kalman filtering techniques that have been proposed for receivers;
e.g., \cite{Komninakis_Sayed_Kalman, GG_Kalman_Rx}. 

%{\color{red}discussion of uncertainty models here. Include the special case of Gaussian models,
%appropriate for TDD and for Tracking case.}

One approach to finding good solutions to outage-based QoS problems 
%for the downlink
%
%Several techniques for finding good linear precoders for such problems have
%been developed~\cite{Chalise_2007_CC_DL,Michael_Asilomar08,Ma_etal_ICASSP2011_arxiv}.
%In addition, ``power loading'' techniques have been developed for cases in which 
%the directions of transmission have already been chosen~\cite{Payaro_2007,Vucic_Boche}.
%
%The principle that underlies the previous approaches  
%to outage-based QoS problems for the downlink \cite{Michael_Asilomar08,Ma_etal_ICASSP2011_arxiv,Payaro_2007,Vucic_Boche} 
is to seek a deterministic approximation of the outage constraint that is conservative and can be represented in a form that is convex in design variables. 
The conservative nature of the approximation 
means that any 
%feasible 
point
\changet{that satisfies the constraints} in the resulting restricted optimization problem will satisfy the original outage constraint\changet{s},
and
the convex nature of approximations in the approaches 
means that a globally optimal solution to the restricted optimization problem can be efficiently found.
%As explained in the next section, the principles that underlie the proposed approaches are somewhat different.
 This approach has led to effective techniques for finding good linear precoders~\cite{Chalise_2007_CC_DL,Michael_Asilomar08,Ken_outage},
 and good ``power loading'' techniques for cases in which 
the directions of transmission have already been chosen~\cite{Payaro_2007,Vucic_Boche}.

The proposed approach is somewhat different in that it 
does not involve an approximation of the outage
constraint, but employs a precise deterministic representation for the case of 
Gaussian uncertainties~\cite{Raphaeli,Hassibi_ISIT_2009}. Unlike the
previous approaches, the resulting optimization problem is not convex, but 
%\revtsix{based on connections with the framework of interference functions~\cite{Yates_interference_functions}},
we develop  a straightforward 
cyclic coordinate descent algorithm that,
\revtsev{through connections with the framework of interference functions~\cite{Yates_interference_functions},
can be shown to converge to an optimal solution when the starting point \changet{satisfies the outage constraints}.}
\changet{(A related fixed-point algorithm was developed concurrently in \cite{Grundinger}.)}
%typically produces good solutions. Indeed, in a number of scenarios
\revtsix{Even when terminated quite early, this algorithm typically provides}
%our suboptimal solutions to the precise formulation of the problem provide 
superior performance to that of 
the globally optimal solutions to the conservative approximation.

%In Section~\ref{sec:ZF} we use insight 
Insight developed from our initial implementation of the basic principle is then used
%from that development
 to construct a more computationally efficient  power-loading technique for
the case of nominally ``zero-forcing'' beamforming directions. While that technique does involve a conservative approximation, the structure of the approximation is quite different from those that have been previously applied,
and  
numerical experience suggests that it can be significantly less conservative. % than some the existing techniques.
Interestingly, in some important scenarios the lower level of conservatism in the approximation means that
the proposed power loading algorithm with nominally zero-forcing   directions yields 
better performance than existing techniques in which the power loading and   directions are designed jointly.

\changet{The remainder of the paper is organized as follows: The system model for the downlink
and the uncertainty model that we will consider are described in Section~\ref{sec:model}. Having established
those models, in Sections~\ref{sec:cc_robust_precoding} and \ref{sec:cc_robust_pl}, we formally define the problems of robust precoding and robust power
loading, respectively. We also review some of the existing approaches to those problems. In Section~\ref{sec:closed_form} we provide
the result from \cite{Hassibi_ISIT_2009} that enables us to write closed-form deterministic expressions for the outage probability (in the case
of Gaussian uncertainties). In Section~\ref{sec:general_algo} we present a  coordinate descent algorithm for optimal power loading in the case of a generic selection of the beamforming directions. In Section~\ref{sec:ZF} we present several tailored algorithms for the case of the zero-forcing directions (for the base station's estimates of the channels). These algorithms are based on an approximation of the integrand in the generic method. That approximation enables the integral to be computed using residue techniques, which significantly reduces the computational cost of the algorithm. In Section~\ref{sec:sims} we compare the performance of the proposed algorithms to a number of existing approaches, and conclusions are drawn in
Section~\ref{sec:conc}. Some of the details of the technical results are discussed in the Appendices.}

%======================================================================
%==================System Model===========================================
%=========================================================================
\section{System Model} 
\label{sec:model}

\begin{figure}
\centering
\includegraphics[width=0.9\figwidth]{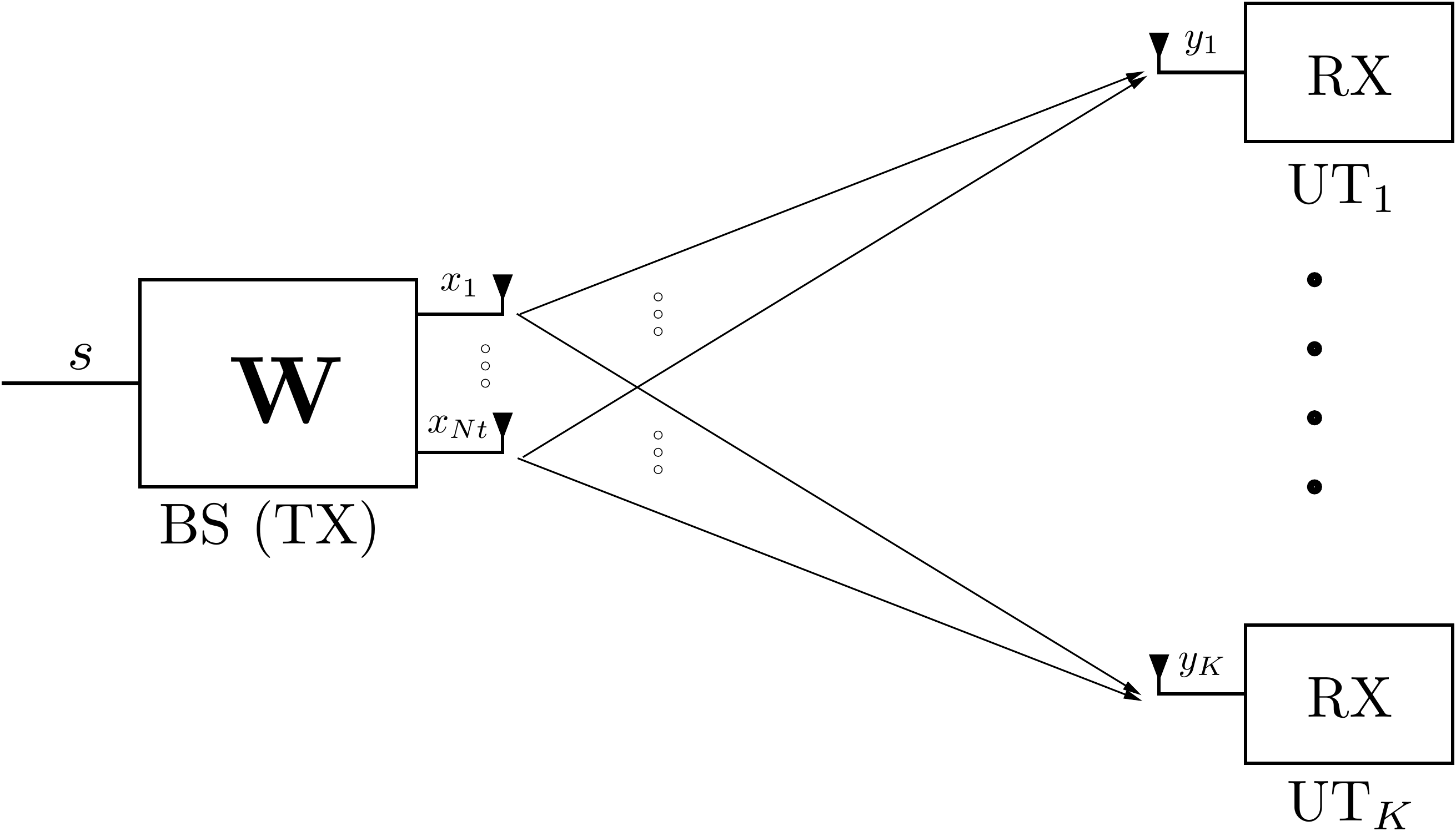}
\caption{A single-cell downlink setting with $N_t$ antennas at the base station (BS) and $K$ users, each with a single antenna.} \label{fig:MIMO}
\end{figure}

We consider a narrowband single-cell downlink scenario in which a base station with ${N_t}$ antennas  
sends independent messages to $K$ users (unicast transmission), each of which is equipped with a single antenna, as illustrated in Fig.~\ref{fig:MIMO}.
%%%%%%%%%% Fig 2-1
The base station employs linear precoding and the transmitted signal at each channel use is 
% 2.1
\begin{equation}
\label{eq:trans_sig}
{\mathbf{x}} = \sum_{k=1}^{K} {\mathbf{w}}_k s_k = {\mathbf{W}}{\mathbf{s}},
\end{equation}
where ${\mathbf{w}}_k \in \mathbb{C}^{N_t}$ is the beamforming vector for the $k^{th}$ user and forms the $k^{th}$ column of 
 the precoding matrix 
${\mathbf{W}} \in \mathbb{C}^{N_t \times K}$, 
and $s_k$ is the symbol to be sent to the $k^{th}$ user.
We normalize the symbols  so that they have unit energy, and since the messages are assumed to be independent we have  $E\{{\mathbf{s}} {\mathbf{s}}^H\} = {\mathbf{I}}$.  
For user $k$, the received signal can be modelled as
\begin{equation}
%2.2
\label{eq:recieved_sig}
y_k = {\mathbf{h}}_k^H \mathbf{x} + {z}_k,
\end{equation}
where ${\mathbf{h}}_k^H \in \mathbb{C}^{N_t}$ is 
the row vector of complex channel gains from 
the transmitting antennas 
to the $k$\textsuperscript{th} receiver, and $z_k$ denotes additive noise, which is assumed to be circular complex Gaussian with zero mean and variance $\sigma_k^2$. The received signal can be rewritten as
%2.3
\begin{equation}
\label{eq:recieved_sig_version2}
y_k = {\mathbf{h}}_k^H \mathbf{w}_k s_k +  {\mathbf{h}}_k^H \bar{\mathbf{W}}_k \mathbf{s}+ z_k,
\end{equation}
where $\bar{\mathbf{W}}_k = [\mathbf{w}_1,\dots,\mathbf{w}_{k-1},\mathbf{0},\mathbf{w}_{k+1},\dots,\mathbf{w}_K]$. %Assuming that each receiver has perfect knowledge of %the gain of intended signal, 
%${\mathbf{h}}_k^H \mathbf{w}_k$, the
The first term in \eqref{eq:recieved_sig_version2} is the useful signal term for coherent detection of the message sent to user $k$, while the second term represents the interference due to the transmissions to the other users and the third term is the noise. 

The desire to provide \changet{fixed-rate} service\changet{s} to several classes of users
%, and the desire to support such services with low latency requirements, 
has led to development of design techniques which guarantee that a certain quality-of-service (QoS) constraint is satisfied for each user. 
%In this paper, the quality of service is specified in terms of a measure of the signal-to-interference-and-noise ratio (SINR) at user $k$.
%The appropriate measure of the SINR is dependent on the structure of the receiver and the available information at each user terminal. We will consider 
In the case of coherent single-user detection, in which 
%each user has perfect knowledge of ${\mathbf{h}}_k^H \mathbf{w}_k$ and 
interference is treated as noise, 
%and we will assume that . In that setting, 
the QoS is typically specified in terms  of the SINR at the $k^{th}$ user, 
%2.4
\begin{equation}
\label{eq:SINR_def_NEW}
\mathsf{SINR}_k(\mathbf{W}) =\frac { |\mathbf{h}_k^H {{\mathbf w}_k}|^2} {\mathbf{h}_k^H{\mathbf {\bar W}}_k {\mathbf {\bar W}}^H_k \mathbf{h}_k+{\sigma}_k^2}. 
\end{equation}

If the BS has perfect knowledge of the channels, ${\mathbf{h}}^H_1,\dots,{\mathbf{h}}^H_K$, then for any choice of the beamforming matrix $\mathbf{W}$, it can compute each receiver's SINR and hence it can adapt its transmission. In a standard scenario, the operator may wish to have the BS adapt its transmission so that it minimizes
the power,  $E\{\mathbf{x}^H\mathbf{x}\} = \sum_k \|\mathbf{w}_k\|^2 = \Trace{(\mathbf{W}\mathbf{W}^H)}$, required to provide selected users with a specified target SINR, or declare that it is not possible to meet that specification. That is, the BS may be required to find a set of beamformers $\{\mathbf{w}_k\}$ that solves
%2.5
\begin{subequations}
\label{eq:perfect_CSI_problem}
\begin{align}
\min_{\{\mathbf{w}_k \in \mathbb{C}^{N_t}\}_{k=1}^K} &\quad \Trace \bigl({\mathbf{WW}}^H\bigr) \\
\text{subject to}&\quad  \mathsf{SINR}_k(\mathbf{W}) \geq  {\gamma}_k,  \quad \quad \forall k=1,\dots,K, 
\label{eq:perfect_CSI_problem_constraints}
\end{align}
\end{subequations}
or show that no such set of beamformers exists. Here, $\gamma_k$ is the specified SINR for user $k$. This problem can be efficiently solved by 
transforming it into a convex second order cone program \cite{Wiesel_2006_Fixed};
\newchanget{by applying a fixed-point mapping~\cite{Rashid-Farrokhi_1998_COM};
by applying the (related) notions of `uplink-downlink' or Langrangian duality \cite{Schubert_Boche_2004,Bjornsson_etal_2014}};
 or by relaxing it to a convex semidefinite program and showing that the relaxation is tight \cite{Beamforming_Bengtsson_2001}.
A `dedicated' training phase is then employed so that the $k^{th}$ receiver can estimate the scalar
$\mathbf{h}_k^H\mathbf{w}_k$ with an accuracy that is sufficient to perform the coherent detection
that is implicit in \eqref{eq:SINR_def_NEW}.

In practice, the CSI that can be made available at the BS is imperfect, due to estimation errors,
quantization, delays, and other effects; e.g., \cite{Caire_Jindal_Kobayashi_Ravindran,Xu_Andrews_Jafar},
and yet the design of the downlink precoder with QoS targets remains an important problem. 
As simple strategy is to perform a ``mismatched'' design in which the BS's estimates of the channels are
treated as if they were accurate, and the SINR targets $\gamma_k$ are increased in an attempt to
increase the probability that the original targets are satisfied for the actual channels.
%{\color{red} should we simulate this model?}
%
%In practice, most wireless communication environments are dynamic, in the sense that the positions of the user terminals and/or the scattering environment change over time. In such settings, providing the BS with highly accurate estimates of the channels to the users may require a significant fraction of the available communication resources.
%Therefore, the BS must perform the design with imprecise knowledge of the channels. 
%%
%%{\color{red}phases of communication here}
%%
%There are a number of different strategies that can be taken in pursuing that design. The simplest strategy is to perform the design as if the estimates were exact and to hope that, over time, there is a reasonable likelihood that the performance requirements are met. That approach can be effective in terms of the sum-rate degrees-of-freedom
%in applications with low mobility~\cite{Xu_Andrews_Jafar}, but it is less effective for the QoS problem in~\eqref{eq:perfect_CSI_problem}.
%%{\color{red} effective in terms of sum-rate DoF (Jafar Andrews) but not necess for problems of this form?}
%A variation on that strategy is to perform the design in the ``mismatched'' manner, but to increase the SINR targets $\gamma_k$ beyond the required level in an attempt to increase the likelihood that the mismatched design yields a solution that meets the original requirements.
%
An alternative strategy is to construct a model for the uncertainty in the BS's knowledge of the channel and to take that uncertainty into account in the formulation of the design problem. The model for the uncertainty is dependent on the method by which the BS obtains information about the state of the channel. 

In this paper, we will consider scenarios in which a Gaussian model for that uncertainty is appropriate.
\newchanget{Among several scenarios in which  such a model is appropriate, a prominent one}
%A prominent example 
is the case of (single-cell) time-division duplex (TDD) systems, in which the base station 
\newchanget{obtains an estimate of each}
%estimates
%the 
channel via training and 
%{\color{red} \Large \bfseries linear MMSE} 
\newchanget{linear}
estimation on the uplink \newchanget{(e.g., \cite{Biguesh_Gershman_training})}, using
the principle of reciprocity of the channel~\cite{smith2004direct}. In that setting, if the 
channel changes sufficiently slowly 
\newchanget{(with respect to the ``ping-pong time'' of the TDD system)}, 
and \newchanget{if} appropriate RF calibration is performed (e.g., \cite{kaltenberger2010relative}),
\newchanget{then the estimation error is the dominant source of the mismatch between the actual and estimated channels.}
\newchanget{When the linear estimator is unbiased}, 
the  base station's estimate of the (baseband equivalent) channels can be written as
\begin{equation}
\hat{\mathbf{h}}_k^H = \mathbf{h}_k^H +\mathbf{e}_k^H,
\label{eq:add_uncert}
\end{equation}
where $\mathbf{e}_k$ is an zero-mean Gaussian random variable that is independent of $\mathbf{h}_k$
and has a covariance matrix, $\mathbf{C}_k$.
%, that is a function 
%of the 
%channel covariance, 
%and 
%the structure and SNR of the training sequence, 
%\newchanget{and, in some cases, the covariance of the channel}; e.g., \cite{Poor:1994:ISD:174518}.
That is, $\mathbf{e}_k$ is Gaussian with $E\{\mathbf{e}_k\}=\boldsymbol{0}$ and
$E\{\mathbf{e}_k\mathbf{e}_j^H\}=\delta[k-j]\mathbf{C}_k$, where $\delta[\cdot]$ denotes
the Kronecker delta. 
\newchangem{In particular, if the noise on the uplink is circular complex Gaussian with zero mean
and covariance $\sigma_{\text{BS}}^2\mathbf{I}$, then if we treat the channel vector as being deterministic,
the best linear unbiased estimator (BLUE)
is actually the least-squares estimator and the covariance matrices 
$\mathbf{C}_k$ depend on
the structure and SNR of the training sequence; e.g., \cite{Biguesh_Gershman_training,Poor:1994:ISD:174518}.
In this setting, the  least-squares estimator is also a minimum variance unbiased estimator. 
In the case of orthogonal training, which is optimal in this setting~\cite{Biguesh_Gershman_training}, the
covariance matrices reduce to $\mathbf{C}_k=\sigma_e^2\mathbf{I}$, where $\sigma_e^2$ is a function of the training SNR and the 
length of the training sequence.}

\section{Design of Outage-based Robust Downlink Transmitters}
\label{sub:BF_selection}
%%%%%%%%%%%%%%%%%%%%%%%%%%%%%%%%%%%%%%%%%%%%%%% 

The focus of this paper is on robust power loading schemes for systems in which the beamforming directions
$\mathbf{w}_k/\|\mathbf{w}_k\|$ have been chosen; see Section~\ref{sec:cc_robust_pl}. 
However, for completeness we first
consider the case of full precoder design.

\subsection{Chance-constrained Robust Precoding}
%%%%%%%%%%%%%%%%%%%%%%%%%%%%%%%%%%%%%%%%%%%%%%% 
\label{sec:cc_robust_precoding}
For given SINR targets $\gamma_k$, the quality of service constraint   is that the probability that
${\mathsf{SINR}_k(\mathbf{W}) \geq  {\gamma}_k}$ should be greater than $1 - \epsilon_k$, for a pre-specified ``probability of outage'' $\epsilon_k$. Therefore, given the uncertainty model $\mathbf{h}_k = \hat{\mathbf{h}}_k+\mathbf{e}_k$ and a distribution for $\mathbf{e}_k$, the problem of interest can be written as
%2.17
\begin{subequations}
\label{eq:original_problem}
\begin{align}
\min_{\{\mathbf{w}_k \in \mathbb{C}^{N_t}\}_{k=1}^K} &\quad \Trace \bigl({\mathbf{WW}}^H\bigr) \\
\text{subject to}&\quad  \operatorname{Pr}_{{\mathbf e}_k} \bigl(\mathsf{SINR}_k(\mathbf{W}) \geq  {\gamma}_k\bigr) \geq 1 - \epsilon_k,  \quad \quad \forall k, 
\label{eq:original_chance_constraints}
\end{align}
\end{subequations}
where the SINR at user $k$ was defined   in \eqref{eq:SINR_def_NEW}. The presence of the chance constraints in 
\eqref{eq:original_chance_constraints}
 makes  the problem difficult to tackle directly; 
 %especially because 
 the SINR in \eqref{eq:SINR_def_NEW} is the ratio of   quadratic functions of the design variables. One approach is to apply a conservative transformation to the SINR constraint in \eqref{eq:original_chance_constraints} to convert the problem in \eqref{eq:original_problem} into a chance-constrained second-order cone program (SOCP)~\cite{Michael_Asilomar08}. By applying various conservative 
 approximations of chance-constrained SOCPs,  efficiently-solvable
 deterministic convex optimization problems are obtained (some are SOCPs, others are semidefinite programs, SDPs). 
 The  conservative nature of the approximations means that when these convex problems are feasible,%
 \footnote{\changet{in the sense that there exists at least one choice for the set of beamformers $\{\mathbf{w}_k\}_{k=1}^K$ that satisfies all the constraints in the conservative approximation}} the  solution is guaranteed to satisfy the chance constraints in \eqref{eq:original_chance_constraints}. 
 
 The related approach of \cite{Ken_outage} first applies a semidefinite relaxation to the problem in 
 \eqref{eq:original_problem}, which yields a semidefinite program (SDP) with chance constraints on quadratic functions of a vector of variables. These chance constraints are then conservatively approximated by deterministic 
 convex constraints leading to an SDP formulation. 
 The solution to that SDP formulation is guaranteed to satisfy the original chance constraints in \eqref{eq:original_chance_constraints} whenever each of the solution matrices 
 %in \eqref{eq:cons_precoder_RARstyle} 
 has rank one.
Numerical experiments    suggest that this is almost always the case,
especially when the conservative approximation of the chance constraint results in a spherical
uncertainty region, as distinct from the more general elliptical case \cite{Ken_outage,Luo_SDR_tight}. 
%{\color{red} add the citation of Tom's EURASIP paper here.}
Under the zero-mean Gaussian uncertainty model in \eqref{eq:add_uncert}, one 
  of the SDP problem formulations obtained using this approach 
  %for a zero mean i.i.d.\ Gaussian uncertainty model 
  involves optimizing over
 Hermitian matrices $\mathbf{U}_k$ that represent $\mathbf{w}_k\mathbf{w}^H_k$, but are not
 required to be rank~1,
 %where $\mathbf{U}_k$ is a Hermitian matrix, $\mathbf{U}_k \in \mathbb{H}^{N_t \times N_t}$, 
 and conservatively approximating the chance constraint by a linear matrix inequality \cite{Ken_outage}. The resulting SDP is
 %2.18
\begin{subequations}
\label{eq:cons_precoder_RARstyle}
\begin{align}
\min_{\{\mathbf{U}_k \in \mathbb{H}^{N_t \times N_t}\}\,, \{t_k\geq0\}} & \quad 
\sum_k \Trace \bigl(\mathbf{U}_k \bigr) \\ 
\text{subject to}&\quad \begin{bmatrix}
{\mathbf Q}_k+t_k{\mathbf I} & {\mathbf r}_k \\
{\mathbf r}^H_k & v_k-{t_k}{d_k}^2 
\end{bmatrix} \succeq {\mathbf 0}, \quad \quad \forall k,
\end{align}
\end{subequations}
 where $\mathbf{Q}_k = \mathbf{C}^{1/2}_k\Bigl( \frac{1}{\gamma_k} \mathbf{U}_k - \sum_{j\not=k}  \mathbf{U}_j \Bigr) \mathbf{C}^{1/2}_k$, $\mathbf{r}_k = \mathbf{C}^{1/2}_k\Bigl( \frac{1}{\gamma_k} \mathbf{U}_k - \sum_{j\not=k}  \mathbf{U}_j \Bigr)\hat{\mathbf{h}}_k$, $v_k = \hat{\mathbf{h}}_k\Bigl( \frac{1}{\gamma_k} \mathbf{U}_k - \sum_{j\not=k}  \mathbf{U}_j \Bigr)\hat{\mathbf{h}}_k-\sigma^2_k$ and $d_k = {\sqrt {{\phi^{-1}}_{{X}^2_{2N_t}} (1-\epsilon_k)/2}}$, where ${\phi^{-1}}_{{X}^2_{2N_t}}(\cdot)$ is the inverse cumulative distribution function of central Chi-square random variable with $2N_t$ degrees of freedom. We will use the formulation in \eqref{eq:cons_precoder_RARstyle}
 as a benchmark in the evaluation of the proposed designs.

%%%%%%%%%%%%%%%%%%%%%%%%%%%%%%%%%%%%%%%%%%%%%%%
%					Chance-constrained robust power loading					%
%%%%%%%%%%%%%%%%%%%%%%%%%%%%%%%%%%%%%%%%%%%%%%%
 \subsection{Chance-constrained Robust Power Loading}
 \label{sec:cc_robust_pl}
%%%%%%%%%%%%%%%%%%%%%%%%%%%%%%%%%%%%%%%%%%%%%%%
In robust precoding
the directions of transmission, ${\breve{\mathbf {w}}}_k = {{\mathbf w}_k}/{{||{\mathbf w}_k||}_{2}}$, and the power allocated to each direction, $\breve{p}_k = ||{\mathbf w}_k||^2_2$,
are found jointly. A potentially simpler approach is to choose the directions $\breve{\mathbf{w}}_k$ based on the transmitters' channel estimates $\hat{\mathbf {h}}_k$ and then to seek solutions to the problem in 
\eqref{eq:original_problem}
 over the $K$ powers, $\breve{p}_k$. 
It is often more convenient to remove the restriction that the 
 directions be specified in a normalized form, and
 simply pre-specify vectors. We will pre-specify the directions as not necessarily normalized vectors 
 $\mathbf{b}_k$ and  seek a power allocation $\{p_k\}_{\changet{k=1}}^{\changet{K}}$ for these vectors so that ${\mathbf w}_k = \sqrt{p_k}\, {\mathbf b}_k$. In that case, the 
total power transmitted is $\sum_{k=1}^{K} p_k \| {\mathbf b}_k\|^2_2$. If we define ${\mathbf B}= [{\mathbf b}_1,{\mathbf b}_2,...,{\mathbf b}_K ]$ and the diagonal matrix ${\mathbf P} = \operatorname{Diag}(p_1,p_2,...,p_K)$, the robust power loading problem can be written as
%2.19
\begin{subequations}
\label{eq:power_load_original}
\begin{align}
\hspace{-0.5em}\min_{\{p_k\geq0\}}&\; \Trace \bigl({\mathbf{BPB}}^H\bigr)\\
\text{s.t.}&\;
\operatorname{Pr}_{\mathbf{e}_k} \Bigl(\frac { |({{\hat {\mathbf{h}}}_k}^H+{\mathbf{e}}^H_k) {{\mathbf b}_k}|^2 p_k} {({{\hat {\mathbf h}}_k}^H+ {{\mathbf e}}^H_k){\mathbf {\bar B}}_k {\mathbf P} {\mathbf {\bar B}}^H_k (  {{\hat {\mathbf h}}_k}+{\mathbf e}_k)+{\sigma}_k^2}\geq  {\gamma}_k\Bigr)  \notag \\
&\qquad\qquad\qquad\qquad \qquad\qquad \geq 1 - \epsilon_k, \quad \quad \forall k,  
\label{eq:chance_constraint_for_original_power_load}
\end{align}
\end{subequations}
where ${\bar{\mathbf{B}}}_k = [{\mathbf b}_1,...,{{\mathbf b}_{k-1}, {\boldsymbol{0}},{\mathbf b}_{k+1},...,{\mathbf b}_K}]$.
 A common choice for the precoding matrix ${\mathbf B}$ is the regularized channel inversion 
 precoder for the estimated channel \cite{Swindle}: Given matrix of channel estimates 
${\hat {\mathbf{H}}} = [{\hat {\mathbf{h}}}_1,{\hat {\mathbf{h}}}_2,...,{\hat {\mathbf{h}}}_K]^H$ and a non-negative real number $\alpha$,
 %2.20
\begin{equation}
\label{eq:RCI_Beamformer}
{\mathbf{B}}_{\text{RCI}} = {\hat {\mathbf{H}}}^{H} \bigl({\hat {\mathbf{H}}} {\hat{\mathbf {H}}}^H+{\alpha {\mathbf I}_K}\bigr)^{-1}.
\end{equation}
In the special case when $\alpha=0$ \newchanget{(and $K\leq N_t$)}, the nominal zero-forcing precoder is obtained, % where 
%2.21
\begin{equation}
{\mathbf{B}}_{\text{ZF}} = {\hat {\mathbf{H}}}^{H} \bigl({\hat {\mathbf{H}}} {\hat{\mathbf {H}}}^H\bigr)^{-1}.
\end{equation}

In the development of approaches to solve the problem in \eqref{eq:power_load_original}, we will rewrite the chance constraints in \eqref{eq:chance_constraint_for_original_power_load} in the form of chance constraints on a quadratic function of a standard complex Gaussian random variable, 
$\boldsymbol{\delta}_k\sim CN(\boldsymbol{0},\mathbf{I})$, namely, 
%2.22
\begin{align}
\operatorname{Pr}_{\boldsymbol{\delta}_k} \bigl(\boldsymbol{\delta}^H_k {\mathbf Q}_k\boldsymbol{\delta}_k+2\operatorname{Re}(\boldsymbol{\delta}^H_k {\mathbf r}_k) + v_k  \geq  0\bigr) \geq 1 - \epsilon_k, 
\label{eq:delta_probability}
\end{align}
where
\begin{math}
{\mathbf Q}_k  = {{\mathbf C}_k}^{1/2}\bigl( {\frac {p_k}{\gamma_k}} {\mathbf b}_k {\mathbf b}^H_k - {{\mathbf {\bar B}}_k} {\mathbf P}{{\mathbf {\bar B}}_k}^H \bigr) {{\mathbf C}_k}^{1/2}
\end{math},
\begin{math}
{\mathbf r}_k  = {{\mathbf C}_k}^{1/2}  \bigl( {\frac {p_k}{\gamma_k}} {\mathbf b}_k {\mathbf b}^H_k - {{\mathbf {\bar B}}_k} {\mathbf P}{{\mathbf {\bar B}}_k}^H \bigr) {{\hat {\mathbf{h}}}_k}\end{math}
and 
\begin{math}v_k = {{\hat {\mathbf {h}}}_k}^H \bigl( {\frac {p_k}{\gamma_k}} {\mathbf b}_k {\mathbf b}^H_k - {{\mathbf {\bar B}}_k} {\mathbf P}{{\mathbf {\bar B}}_k}^H \bigr) {{\hat {\mathbf {h}}}_k} - {\sigma}_k^2.
\end{math}
By writing the chance constraints in this form, a number of existing conservative deterministic approximations to the chance constraint can be applied in a straightforward way; see \cite{Ken_outage}. For example, given ${\mathbf B}$, the solution to the following SDP yields powers $\{p_k\}$ that satisfy the constraints in \eqref{eq:power_load_original},
%2.23
\begin{subequations}
\label{eq:cons_powerload_RARstyle}
\begin{align}
\min_{\{p_k\geq0\},\, \{t_k\geq0\}} & \quad 
\Trace \bigl({\mathbf{BPB}}^H\bigr) \\ 
\text{subject to}&\quad \begin{bmatrix}
{\mathbf Q}_k+t_k{\mathbf I} & {\mathbf r}_k \\
{\mathbf r}^H_k & v_k-{t_k}{d_k}^2 
\end{bmatrix} \succeq {\mathbf 0}, \quad \quad \forall k,
\end{align}
\end{subequations}
where $d_k = {\sqrt {{\phi^{-1}}_{{X}^2_{2N_t}} (1-\epsilon_k)/2}}$.
That said, since \eqref{eq:cons_powerload_RARstyle} is based on a conservative approximation, the absence of a solution to \eqref{eq:cons_powerload_RARstyle} does not necessarily mean that there are no powers that satisfy \eqref{eq:power_load_original}.

The goal of this paper is to propose robust power loading algorithms that reduce conservatism and may reduce the computational cost.   Rather than being  based on seeking tractable convex, but conservative
formulations, the proposed approach is based on a closed-form expression for the probability that the SINR constraint is satisfied. 
\section{A Closed-form Expression for the CDF of a Quadratic Function of a Gaussian Random Vector}
\label{sec:closed_form}
\revtsix{The proposed approaches to the  robust power loading problem in \eqref{eq:power_load_original}
will be based on the following closed-form expression} 
%In this section we present a result from \cite{Hassibi_ISIT_2009} that provides a closed-form expression 
for the cumulative distribution function (CDF) of a quadratic function of a standard circular complex Gaussian random vector~\cite{Hassibi_ISIT_2009}. 
%This result is a key step in the development of the proposed design algorithms. In the statement of that
\revtsix{To state that expression,}
%result and the remainder of the paper 
we will use the notation $\complexunit$ to denote $\sqrt{-1}$, and
%will use 
$\|\mathbf{u}\|_{\mathbf{M}}^2$ to denote $\mathbf{u}^H\mathbf{M}\mathbf{u}$.

\begin{lemma}[\cite{Hassibi_ISIT_2009}]
Given a deterministic Hermitian symmetric matrix $\mathbf{M}$ and a deterministic vector $\mathbf{z}$,
for the standard circular complex Gaussian random vector ${\mathbf x} \sim CN(\boldsymbol{0},{\mathbf I})$
%the CDF of 
  %$\|{\mathbf x} - {\mathbf a}\|_{{\mathbf{Q}}}^2 = ({\mathbf x}-{\mathbf a})^H{\mathbf Q}({\mathbf x}-{\mathbf a})$,
 %$\operatorname{Pr}\bigl(\|{\mathbf x} - {\mathbf a}\|_{{\mathbf{Q}}}^2 \leq \tau\bigr)$,  is
% 3.1
\begin{multline}
\label{eq:hasibi}
\operatorname{Pr}\bigl(\|{\mathbf x} - {\mathbf z}\|_{{\mathbf{M}}}^2 \leq \tau\bigr)\\
={\frac{1}{2 \pi}} \int_{-\infty}^{\infty} \frac{e^{\tau(\complexunit\omega+\beta)}}{\complexunit\omega+\beta} \frac{e^{-c}}{\det({\mathbf I}+(\complexunit\omega+\beta){\mathbf M})}d\omega,
\end{multline}
for some $\beta > 0$ such that ${\mathbf I}+{\beta}{\mathbf M}$ is positive definite. If we let $\mathbf{M}=\mathbf{V}\boldsymbol{\Lambda}\mathbf{V}^H$
denote the eigen decomposition of $\mathbf{M}$, with $\lambda_m$ denoting
the eigenvalues arranged in descending order
$\bigl(\boldsymbol{\Lambda}=\operatorname{Diag}(\lambda_1,\lambda_2,\dots)\bigr)$,
and if we define $\tilde{\mathbf{z}}=\mathbf{V}^H\mathbf{z}$, the constant~$c$ can be written as
\begin{math}
c = \sum_{m=1}^{M} {\frac{|\tilde{z}_m|^2(\complexunit\omega+\beta){\lambda_m}}{1+(\complexunit\omega+\beta){\lambda_m}}}
\end{math}.   
 \label{lem:Hassibi}
\end{lemma}

The second statement in Lemma~\ref{lem:Hassibi} is slightly more general than that in \cite{Hassibi_ISIT_2009} because it does not require $\mathbf{M}$ to be invertible. 
\revtsev{(A complete proof is provided in~\cite{Foad_thesis}.)}
In the application herein, that is important when there are fewer active users than transmitting antennas. 
%As there are some weaknesses in the proof of Lemma~\ref{lem:Hassibi} provided in \cite{Hassibi_ISIT_2009}, a complete proof is provided %
%in~\cite{Foad_thesis}.

\revtsix{As an aside, we observe that alternative approaches to those proposed below can
be developed by considering   circular complex Gaussian random variables $\mathbf{g}_k\sim
CN(\hat{\mathbf{h}}_k,\mathbf{C}_k)$, rewriting the
probability on the left hand side of 
\eqref{eq:chance_constraint_for_original_power_load} as
\begin{equation}
\operatorname{Pr}_{\mathbf{g}_k}\bigl(\mathbf{g}_k^H (
\tfrac{p_k}{\gamma_k}\mathbf{b}_k\mathbf{b}_k - \bar{\mathbf{B}}_k\mathbf{P}\bar{\mathbf{B}}_k^H)
\mathbf{g}_k - \sigma_k^2 \geq 0\bigr) \geq 1-\epsilon_k,
\end{equation}
and employing the closed-form expression for this probability that can be obtained by applying
the residue-based analysis in~\cite{Raphaeli}. While that would be effective for the
generally-applicable algorithm developed in Section~\ref{sec:general_algo}, 
\revtsev{the infinite series in the expression obscure insight. The}  insight that lead
to the development of the tailored algorithms in Section~\ref{sec:ZF} arose from the expression
in Lemma~\ref{lem:Hassibi}.}

\section{Feasible Coordinate Descent Algorithm}
\label{sec:general_algo}
%{\color{red} problem of $i$ and $j$ for indicies and $i$ for complex unit}
%In order to employ Lemma~\ref{lem:Hassibi} in our context, we reformat 
\revtsix{By reformatting} the chance constraint in \eqref{eq:delta_probability} as
%, which is
%%3.2 
%\begin{equation}
%\label{eq:QRV_const}
%\operatorname{Pr} \bigl(\boldsymbol{\delta}^H_k {\mathbf Q}_k\boldsymbol{\delta}_k+2\operatorname{Re}(\boldsymbol{\delta}^H_k {\mathbf r}_k) + v_k  \geq  0\bigr) \geq 1 - \epsilon_k,
%\end{equation} 
%into a format that is compatible to Lemma~\ref{lem:Hassibi}. The resulting expression is
%3.3
\begin{equation}
\operatorname{Pr}\bigl(\|{\boldsymbol{\delta}_k} - \mathbf{a}_k\|_{(-{\mathbf Q}_k)}^2 \leq {\tau}_k\bigr) \geq 1- \epsilon_k,
\label{eq:delta_prob_reformat}
\end{equation}
where ${\mathbf a}_k = - {\mathbf{C}_k}^{-1/2} \hat{\mathbf{h}}_k$ and ${\tau}_k = v_k - {\mathbf a}_k^H {\mathbf Q}_k {\mathbf a}_k 
$,
\revtsix{we can employ  Lemma~\ref{lem:Hassibi} and}
%By doing so, 
%By using the results of Section~\ref{sec:closed_form}, 
%we can 
rewrite the   robust power loading problem in \eqref{eq:power_load_original} in a form that is no longer chance-constrained, but is deterministically constrained:
%3.5
\begin{subequations}
\label{eq:power_load_integral}
\begin{align}
\hspace{-0.5em}\min_{\{p_k\geq0\}}&\quad \Trace \bigl({\mathbf{BPB}}^H\bigr)\\
\text{s.t.}&\quad
{\frac{1}{2 \pi}} \int_{-\infty}^{\infty} \frac{e^{\tau_k(\complexunit\omega+\beta)}}{\complexunit\omega+\beta} \frac{e^{-c_k}}{\det({\mathbf I}-(\complexunit\omega+\beta){\mathbf Q}_k)}d\omega \notag \\
 &\qquad\qquad\qquad\qquad \qquad\qquad\quad \geq 1 - \epsilon_k, \quad\forall k, 
\label{eq:chance_constraint_for_integral_power_load}
\end{align}
\end{subequations}
where $\mathbf{Q}_k$ and $\tau_k$ were defined 
following \eqref{eq:delta_probability} and \eqref{eq:delta_prob_reformat}, respectively, 
%analogous to the definition in \eqref{eq:proof_reformulation} 
and $c_k$ has a format analogous to the format of $c$ in Lemma~\ref{lem:Hassibi}. 
This deterministic problem is not convex.
% but we will now develop a simple algorithm that typically generates good solutions  in practice. 
%
\revtsix{However, since}
the integral in \eqref{eq:chance_constraint_for_integral_power_load} is equivalent to $ \operatorname{Pr}
 \bigl(\mathsf{SINR}_k \geq  {\gamma}_k\bigr)$, we can interpret the behaviour of the integral by looking at the definition of the SINR. For fixed-direction beamformers, $\{\mathbf{b}_k\}_{k=1}^K$, we can rewrite the SINR in \eqref{eq:SINR_def_NEW} in terms of the powers as
 %3.6
 \begin{equation}
 \label{eq:SINR_def_convergence}
{\operatorname{SINR}}_k = \frac { |{\mathbf{h}}^H_k {{\mathbf b}_k}|^2 p_k} {\sum_{j \not =  k} | {\mathbf{h}}^H_k {{\mathbf b}_j}|^2 p_j  +{\sigma}_k^2}.
\end{equation}
\revteight{We can also rewrite the term $\bar{\mathbf{B}}_k\mathbf{P}\bar{\mathbf{B}}_k^H$
that appears in \eqref{eq:chance_constraint_for_original_power_load} and \eqref{eq:delta_probability} as $\sum_{j\neq k}p_j\mathbf{b}_j\mathbf{b}_j^H$.
Using insight from \eqref{eq:SINR_def_convergence} and \eqref{eq:chance_constraint_for_original_power_load}, by making the above substitution in
\eqref{eq:delta_probability}, it can be seen that for  fixed $p_j$, $j\neq k$, the integral in \eqref{eq:chance_constraint_for_integral_power_load} is increasing in $p_k$,
and that for fixed $p_k$, the   integral in \eqref{eq:chance_constraint_for_integral_power_load} is
decreasing in each $p_j$, $j\neq k$. This observation suggests the development of a  cyclic
coordinate descent algorithm (e.g., \cite[Sec.~2.7]{Bertsekas_NLP}) in which we start from a 
\changet{power allocation $\{p_k\}_{k=1}^K$ that forms a feasible point%
\footnote{\changet{That is, a power allocation $\{p_k\}_{k=1}^K$ for which all $K$ constraints in \eqref{eq:chance_constraint_for_integral_power_load} are satisfied.}}
 for the problem in
\eqref{eq:power_load_integral},} 
%
%feasible set of powers 
and at the $k^{th}$ step
of the $i^{th}$ cycle we seek to reduce the value of $p_k$ given the current values of the other powers, 
\changet{while maintaining feasibility}.
A feature of that approach is that \revtnine{due to} the above-mentioned features of \eqref{eq:delta_probability}, at the $k^{th}$ step
of the $i^{th}$ cycle we need only consider the $k^{th}$ constraint in \eqref{eq:chance_constraint_for_integral_power_load}; decreasing $p_k$ will
not violate any of the other constraints.}
\revtnine{Furthermore, in concurrent work~\cite{Grundinger} it was shown that the 
power allocation problem in \eqref{eq:power_load_original}, and hence the equivalent
problem in \eqref{eq:power_load_integral},
%and indeed that in \eqref{eq:power_load_original}, 
can be viewed in the framework
of standard interference functions~\cite{Yates_interference_functions}.}
%(Related observations were made in~\cite{Vucic_Boche_2009}.)}
%
%
%We will develop a flexible version of such an algorithm below.%
%
%Before we do so, we observe from \eqref{eq:delta_probability} and the substitution 
%$\bar{\mathbf{B}}_k\mathbf{P}\bar{\mathbf{B}}_k^H=\sum_{j\neq k}p_j\mathbf{b}_j\mathbf{b}_j^H$
%that for $0\leq \epsilon_k < 1$, and fixed $p_j$, $j\neq k$, the smallest $p_k$ that will satisfy
 %\eqref{eq:delta_probability} with equality   (i) is positive; (ii) is non-decreasing (and is typically increasing) with the
 %increase of any $p_j$, $j\neq k$; and (iii) if we scale each $p_j$, $j\neq k$, by $\alpha >1$ then the new minimum value
% for $p_k$ will be smaller than $\alpha$ times the original, due to the presence of the noise term
 %in $v_k$ in  \eqref{eq:delta_probability}. As a result, 
 %the power allocation problem in \eqref{eq:power_load_integral},
%and indeed that in \eqref{eq:power_load_original}, can be viewed in the framework
%of standard interference functions~\cite{Yates_interference_functions}.
%(That observation was also made in the concurrent work in
%\cite{Grundinger}, and related observations were made in~\cite{Vucic_Boche_2009}.) 
As a result, if the initial power allocation vector is feasible, and if we solve for the
minimum feasible $p_k$ at each step 
\revtten{(in which case all the constraints in \eqref{eq:chance_constraint_for_integral_power_load} are satisfied with equality),} 
\revtnine{the} cyclic coordinate descent algorithm
\revtnine{described above} will converge a globally
optimal solution~\cite{Yates_interference_functions,Schubert_Boche_2007}. 
The particular algorithm that we will develop below involves some additional parameters that
will facilitate computational tradeoffs in the algorithm. Although we will describe the algorithm  
%as a cyclic coordinate descent algorithm 
with each step in the cycle in the natural order,
the  principles apply to other orderings and even certain asynchronous updating
schemes~\cite{Yates_interference_functions}.
\revtsix{To put these principles into practice, we let $\mathbf{P}^{(0)}$ denote the diagonal matrix containing
the initial feasible power allocation. (We will discuss techniques for finding such a ${\mathbf{P}}^{(0)}$ below.)}
%\changet{Given a desired accuracy $\Delta_k$ for the $k^{th}$ outage constraint, at} 
At
% into practice in a cyclic coordinate algorithm, at
 the $k^{th}$ step of the $i^{th}$ cycle we
  choose a value for $\Delta^{(i)}_k \leq \Delta^{(i-1)}_k$ and 
 perform \changet{a} bisection search  on the interval $[0,{p}^{(i-1)}_k]$ for a value of ${p}_k$ such that the probability that ${\mathsf{SINR}}_k \geq \gamma_k$ lies in the interval 
% $[1-\epsilon_k,1-\epsilon_k+\changet{\Delta_k}]$
  $[1-\epsilon_k,1-\epsilon_k+\Delta^{(i)}_k]$;
i.e.,
\begin{multline}
%\{{{p}}^{(i)}_k \in [0,{{p}}^{(i-1)}_k] |  
1 - \epsilon_k \leq {\frac{1}{2 \pi}} \int_{-\infty}^{\infty} \frac{e^{\tau_k(\complexunit\omega+\beta)}}{\complexunit\omega+\beta} \frac{e^{-c_k}}{\det({\mathbf I}-(\complexunit\omega+\beta){\mathbf Q}_k)}d\omega  \\ \leq 1 - \epsilon_k +  \Delta^{(i)}_k 
%\changet{\Delta_k}
\label{eq:bisect_interval}
\end{multline}
 %The probability is calculated using the integral expression in \eqref{eq:chance_constraint_for_integral_power_load} 
 with $c_k$, $\tau_k$, $\mathbf{Q}_k$ being calculated using $p^{(i)}_1, \dots, p^{(i)}_{k-1}, p^{(i-1)}_{k+1}, \dots, p^{(i-1)}_{K}$ and the midpoint for the current interval in the bisection search for $p^{(i)}_{k}$. 

The algorithm is terminated once we find a power allocation $\mathbf{P}^{(i)}$ such that for each $k$ the probability that   ${\mathsf{SINR}}_k \geq \gamma_k$ lies in the interval $[1-\epsilon_k,1-\epsilon_k+
\Delta^{\text{min}}_k
%\changet{\Delta_k}
]$, where $\Delta^{\text{min}}_k$ is a pre-specified bound.
\changet{(As is implicit in \eqref{eq:bisect_interval}, the cycle at which the algorithm terminates 
is the first one for  which $\Delta_k^{(i)} \leq  \Delta^{\text{min}}_k$ for all $k$.)}
%, or a pre-specified number of cycles is reached. 
%\revtsev{(Early termination based on the number of cycles can be enabled, if desired.) 
%\textbf{\itshape We can have this parenthetical statement, or we might choose to remove it.}}
A feature of this algorithm is that at each step in each cycle the
power allocation $\{{{p}}^{(i)}_1,{{p}}^{(i)}_2,\dots,{{p}}^{(i)}_k,{{p}}^{(i-1)}_{k+1},\dots,{{p}}^{(i-1)}_K\}$ is feasible and hence whenever the algorithm is terminated, the current power allocation will satisfy the specified QoS constraints of the original problem. Furthermore, at each step in each cycle, the objective value decreases, or remains the same. The latter case occurs when the termination criteria is satisfied prior to performing the current coordinate descent step and therefore the algorithm will go on to the next step without any changes in the power allocation.

%{\color{red}too much in this paragraph}
The parameter %$\changet{\Delta_k}$
 $\Delta^{\text{min}}_k$ 
%is one of the parameters of the algorithm that 
enables  us to make \revtsev{tradeoffs} between the performance and the computational cost of the algorithm. A smaller value for 
$\Delta^{\text{min}}_k$ 
% \changet{\Delta_k}$
results in less conservative solutions that are achieved using less transmitted power,
but will
% However, a smaller value for $\Delta^{\text{min}}_k$ also results in  
%smaller interval for the stopping criteria, which would 
typically lead to a \changet{larger value for the total number of  bisection steps 
(cf. \eqref{eq:bisect_interval}) required for the algorithm to terminate.}
\changett{While the choice of $\Delta^{\text{min}}_k$ controls the performance of
the algorithm and has the dominant influence on its computational cost, the flexibility to choose}
%Hence, for smaller $\Delta^{\text{min}}_k$ the complexity of the algorithm is increased. 
% The other parameter that effects   the computational cost
% % of the algorithm 
% is $\Delta^{(i)}_k$. A smaller value for $\Delta^{(i)}_k$ results in more bisection  steps in the $k^{th}$ step of the $i^{th}$ cycle, and that increases the computational cost of each cycle, but it also produces a better power allocation at the end of each cycle. 
% %That may help to reduce the number of cycles; hence, the nett effect of $\Delta^{(i)}_k$ on the complexity is not explicit. 
%One useful strategy for defining $\Delta^{(i)}_k$ is to pick larger values for early cycles and then choose smaller values as the cycle index increases; i.e., $\Delta^{(i-1)}_k \geq\Delta^{(i)}_k$. Arguably, the simplest strategy is to define $\Delta^{(i)}_k$ to be a constant value equal to $\Delta^{\text{min}}_k$ for all cycles; i.e., $\Delta^{(i)}_k =  \Delta^{\text{min}}_k$, and that is what we will do in the experiments in Section~\ref{sec:sims}.
\changet{%The flexibility of choosing 
$\Delta_k^{(i)}$ at each cycle offers the opportunity to trade  the number of 
bisection steps required in each cycle   against the number of cycles required. 
%IntuitionAlthough insight into effective choices for $\Delta_k^{(i)}$ has proven difficult to extract, 
\changett{A} variety of intuitively motivated decreasing sequences for $\{\Delta_k^{(i)}\}_i$ could be considered, \changett{but}
our numerical experience suggests that the simple choice of $\Delta_k^{(i)}=\Delta_k^{\text{min}}$, which results in a solution being obtained
in a single cycle,} \changett{leads to an effective implementation.}
% is a remarkably effective choice.}
%{\color{red}that sentence will need refinement}
%and this has the potential to  
%\changet{If the resulting tradeoff does not have the desired features, one could specify a sequence of accuracy
%parameters $\Delta_k^{(i)}$ that decreases with the cycle index~$i$ until the desired accuracy, $\Delta_k$, is
%obtained~\cite{Foad_thesis}. However, in our numerical experiments in Section~\ref{sec:sims} we will simply use the
%constant value $\Delta_k$.}

To complete the description of the algorithm, we need to establish a method to determine a feasible starting point,
\revtsix{$\mathbf{P}^{(0)}$}. As the feasible set%
\footnote{\changet{That is, the set of all feasible power allocations}}
 in \eqref{eq:power_load_integral} is not necessarily convex, determining whether or not an instance of the problem in \eqref{eq:power_load_integral} is feasible can be computationally demanding task; \changet{e.g., \cite[Section~5.1]{BT_E_N_Robust_Opt}}. 
%{\color{red}BenTal et al, Robust, Section 5.1}.
\newchanget{Therefore, rather than trying to determine, precisely, whether or not the problem is feasible, we
seek an approach that is computationally cheap and}
%Instead, we simply seek an approach that 
often finds feasible points for reasonable instances of the problem.
The proposed approach involves selecting an initial diagonal power allocation matrix and evaluating 
each of the
integral in \eqref{eq:chance_constraint_for_integral_power_load}.
 If that power allocation is not feasible, the allocation is iteratively doubled until a feasible allocation is found or the power become unreasonably large. In the latter case a new initial power allocation can be selected and the search for a feasible allocation repeated, or the algorithm reports that no feasible point was found. 
 \newchanget{The question that remains is how to choose the initial power allocation.}
 In our experiments we have found that choosing the initial power allocation to be the power allocation that would be chosen if the channel estimates ${\hat{\mathbf{h}}}_k^H$ were exact (perfect CSI) and if each 
$\mathsf{SINR}_k$ were set to be equal to its \newchanget{target value, $\gamma_k$,} typically leads to a feasible starting 
point for the main algorithm
after a small number of \changet{the} doubling iterations \changet{described above}. 
That initial allocation is the solution of the following set of linear equations~\cite{Foad_thesis}:
%{\color{red} that sentence is a bit convoluted. can we shorten this?}
%By rearranging the terms in the SINR expression, one can obtain the initial power allocation by solving a set of linear equations, as we will now show.
%
%By assuming the perfect channel state information, %which means ${\mathbf{e}}_k = \mathbf{0}, \forall k$, 
%the original probabilistic constraint, $\operatorname{Pr} \bigl(\boldsymbol{\delta}^H_k {\mathbf Q}_k\boldsymbol{\delta}_k+2\operatorname{Re}(\boldsymbol{\delta}^H_k {\mathbf r}_k) + v_k  \geq  0\bigr) \geq 1 - \epsilon_k$, is simplified to $v_k \geq 0$, which is a deterministic constraint. According to definition of $v_k$ in \eqref{eq:delta_probability}, 
%%we have \begin{math}v_k = {{\hat {\mathbf {h}}}_k}^H \bigl( {\frac {p_k}{\gamma_k}} {\mathbf b}_k {\mathbf b}^H_k - {{\mathbf {\bar B}}_k} {\mathbf P}{{\mathbf {\bar B}}_k}^H \bigr) {{\hat {\mathbf {h}}}_k} - {\sigma}_k^2.
%%\end{math} We 
%we can rewrite these inequalities in a matrix format as:
%%3.7
\begin{equation}
\label{eq:Linear_eq}
\left( \begin{array}{cccc}
n^2_1 		 & -m^2_{12} 		& \dots 	&  -m^2_{1K}\\
-m^2_{21} 	 & n^2_2     		 & \dots 	&  -m^2_{2K}\\
 \vdots 		 & \vdots 		          & \ddots   &\vdots\\ 
-m^2_{K1}   	 & -m^2_{K2}		 & \dots     &  n^2_K \end{array} \right)
 \begin{pmatrix} p_1 \\ p_2 \\ \vdots \\ p_K \end{pmatrix} 
 =
 \begin{pmatrix} \sigma^2_1 \\ \sigma^2_2 \\ \vdots \\ \sigma^2_K \end{pmatrix},
 \end{equation}
 where \revtsev{$m_{ki} = |{{\hat{\mathbf {h}}}_k}^H {{\mathbf b}_i}|$} and $n_k =  {\frac {1}{\sqrt{\gamma_k}}}|{{\hat{\mathbf {h}}}_k}^H {{\mathbf b}_k}|$.
% {\color{red}\bfseries\itshape please check indexing}
% The desired powers can then be obtained by solving the set of linear equations that arise in the case of equality in \eqref{eq:Linear_eq}.
  \newchanget{A closely related alternative would be to set the initial power allocation so that
  the ``perfect CSI SINRs'' are set to a value above $\gamma_k$, say $(1+\Gamma)\gamma_k$, where $\Gamma$
  denotes the (relative) SINR margin; cf.~\cite{K_Boyd_2002,Stridh_etal_2006}. 
  Among other options, one could also consider using the powers
  obtained from \eqref{eq:cons_powerload_RARstyle} as the initial point, but computing those powers incurs a significant computational cost.}

The proposed robust power allocation algorithm is summarized in Algorithm~\ref{Alg:general_case}. 
%{\color{red}\bfseries\itshape please check that this new while-loop form is actually correct}
\revtsix{If the initial power allocation is feasible, and $\Delta_k^{(i)}\rightarrow0$ 
(and the integral is computed precisely)
the algorithm is guaranteed to converge to a global optimum~\cite{Yates_interference_functions,Schubert_Boche_2007}. 
However, our simple algorithm for finding an initial point does not provide any guarantees.
Nevertheless,}
%As the problem in \eqref{eq:power_load_integral} is not convex, the proposed algorithm is not guaranteed to find 
%the globally optimal solution. Indeed, it  is not even guaranteed to find a feasible point when one exists.
%However,  
we 
will demonstrate in Section~\ref{sec:sims} that by tackling the problem directly, without a conservative
approximation, the proposed approach often provides better performance than the existing conservative
approaches. 
Having said that, the repeated requirement to compute an integral of the form in \eqref{eq:chance_constraint_for_integral_power_load} imposes a significant computational burden. (The SDPs that must be solved in the existing conservative approaches, such as those in \eqref{eq:cons_precoder_RARstyle}, also impose a significant computational burden.)
To address this issue, in the following sections we will develop customized variants of the algorithm for the case of the zero-forcing directions.
%; i.e., ${\mathbf{B}} = {\mathbf{B}}_{\text{ZF}} = {\hat {\mathbf{H}}}^{H} \bigl({\hat {\mathbf{H}}} {\hat{\mathbf {H}}}^H\bigr)^{-1}$. 

 %%%%%%%%%%%% Algorithm 1
\begin{algorithm}[t]
\caption{}
\label{Alg:general_case}
\begin{algorithmic}[1]
	\State Given a feasible diagonal power allocation matrix ${\mathbf{P}}= {\mathbf{P}}^{(0)}$,
	and parameters  %\changet{$\Delta_k$},
	 $\Delta^{(0)}_k$, $\Delta^{\text{min}}_k$, 
	\revtsev{set $i=0$}.
	%and $i_{\text{max}}$, \revtsev{\textbf\itshape take out $i_\max$. not needed if globally convergent}
        % \For{$i = 1 \to i_{\text{max}}$}
        \While{\revtsev{$\exists k$ for which \eqref{eq:bisect_interval} 
         with $\Delta^{(i)}_k$ replaced by $\Delta^{\text{min}}_k $ 
        is not satisfied}}
         		\State $i=i+1$
		 \State For all $k$, choose $\Delta^{(i)}_k$ such that $\Delta^{(i)}_k \leq \Delta^{(i-1)}_k$ 
      		 \For{$k= 1 \to K$} 
		      %  \State Update $c_k$, $\tau_k$, $\mathbf{Q}_k$ 
       			\State Using bisection search, find ${{p}}_k ={{p}}^{(i)}_k\in[0,{p}^{(i-1)}_k]$  
			satisfying \eqref{eq:bisect_interval}

         	\EndFor
%         	\If{\eqref{eq:bisect_interval} with $\Delta^{(i)}_k$ replaced by $\Delta^{\text{min}}_k $ is satisfied $\forall k$} %$1 - \epsilon_k \leq {\frac{1}{2 \pi}} \int_{-\infty}^{\infty} \frac{e^{\tau_k(i\omega+\beta)}}{i\omega+\beta} \frac{e^{-c_k}}{\det({\mathbf I}-(i\omega+\beta){\mathbf Q}_k)}d\omega \leq 1 - \epsilon_k + \Delta^{\text{min}}_k $}
%	      		\State STOP
%		\EndIf
       \EndWhile
\end{algorithmic}
\end{algorithm}

\section{Efficient  Algorithms for the Zero-Forcing Case}
\label{sec:ZF}
%In the previous section  we proposed a coordinate descent algorithm for any fixed-direction beamforming. That algorithm involves repeated calculation of integrals in the form of \eqref{eq:hasibi}. Since calculating of these integrals is computationally expensive, in this section we seek to find ways to reduce the computational cost of the algorithm.
%
%In this section, our approach involves restricting attention to nominally zero-forcing beamforming directions. In that case, 
%In the case of 
\newchanget{When} the nominally zero-forcing directions, ${\mathbf{B}} = {\mathbf{B}}_{\text{ZF}} = {\hat {\mathbf{H}}}^{H} \bigl({\hat {\mathbf{H}}} {\hat{\mathbf {H}}}^H\bigr)^{-1}$, 
\newchanget{are chosen,}
\newchanget{which implicitly requires $K\leq N_t$,}
the structure of the integrand in \eqref{eq:chance_constraint_for_integral_power_load} simplifies, and this simplification facilitates an approximation of the integrand that enables straightforward application of residue theory to obtain an analytic expression for the integral.
%\footnote{\revtsix{A residue-based analysis of the general case of 
%the integrand in \eqref{eq:chance_constraint_for_integral_power_load} 
%is available~\cite{Raphaeli}, but the insights behind
%the approximation arose from direct analysis of the integrand in \eqref{eq:chance_constraint_for_integral_power_load}.}}
That analytic expression can be evaluated much more easily than the integral in 
\eqref{eq:chance_constraint_for_integral_power_load}.
%(The approach in this section was inspired in part by some related work in which a slightly different receiver structure was employed \cite{Davidson_Foad_ICASSP2013}.)
%{\color{red} improve the flow above}

%%%%%%%%%%%%%%%%%%%%%%%%%%%%%%%%%%%%%%%%%%%%%%%%%
%									Problem Formulation									
%%%%%%%%%%%%%%%%%%%%%%%%%%%%%%%%%%%%%%%%%%%%%%%%%
\subsection{Basic formulation for the ZF case}
\label{sec:problem_ZF}
For the case of (nominally) zero-forcing beamforming,  
%${\mathbf{B}}={\mathbf{B}}_{\text{ZF}} = {\hat {\mathbf{H}}}^{H} \bigl({\hat {\mathbf{H}}} {\hat{\mathbf {H}}}^H\bigr)^{-1}$, and hence 
${{\bar{\mathbf{B}}}_k}^H{\hat {\mathbf{h}}}_k = {\mathbf 0}$ and ${{\hat{\mathbf {h}}}_k}^H {{\mathbf b}_k} =1$. 
These simplifications enable us to rewrite the robust power loading problem in \eqref{eq:power_load_original} as
%4.1
%\begin{subequations}
%\label{eq:power_load_original_ZF}
%\begin{align}
%\hspace{-0.5em}\min_{\{p_k\geq0\}}&\; \Trace \bigl(\mathbf{B}_\text{ZF} \mathbf{P} {\mathbf{B}}^H_\text{ZF} \bigr)\\
%%4.1.b
%\text{s.t.}&\;
%\operatorname{Pr}_{\mathbf{e}_k} \Bigl(\frac { |1+{\mathbf{e}}^H_k {{\mathbf b}_k}|^2 p_k} { {{\mathbf e}}^H_k{\mathbf {\bar B}}_k {\mathbf P} {\mathbf {\bar B}}^H_k {\mathbf e}_k +{\sigma}_k^2}\geq  {\gamma}_k\Bigr)
%\geq 1 - \epsilon_k, \quad \quad \forall k,  
%\label{eq:chance_constraint_for_original_power_load_ZF}
%\end{align}
%\end{subequations}
%where ${\mathbf P} = \operatorname{Diag}(p_1,p_2,...,p_K)$, or equivalently, as
%4.2
\begin{subequations}
\label{eq:power_load_original_ZF_b}
\begin{align}
\hspace{-0.5em}\min_{\{p_k\geq0\}}&\; \Trace \bigl(\mathbf{B}_\text{ZF} \mathbf{P} {\mathbf{B}}^H_\text{ZF} \bigr)\\
%4.2.b
\text{s.t.}&\;
\operatorname{Pr}_{\boldsymbol{\delta}_k} \Bigl(\boldsymbol{\delta}^H_k {\mathbf Q}_k\boldsymbol{\delta}_k+({\tfrac {p_k}{\gamma_k}})2\operatorname{Re}(\boldsymbol{\delta}^H_k \tilde{\mathbf {r}}_k) + v_k  \geq  0\Bigr) \notag \\
&\qquad\qquad\qquad\qquad \qquad\qquad\quad\geq 1 - \epsilon_k, \quad \forall k,  
\label{eq:chance_constraint_for_original_power_load_ZF_b}
\end{align}
\end{subequations}
where we have used the form of the chance constraint in \eqref{eq:delta_probability}.
While 
\begin{math}
{\mathbf Q}_k
%  =  {{\mathbf C}_k}^{1/2}\bigl( {\frac {p_k}{\gamma_k}} {\mathbf b}_k {\mathbf b}^H_k - {{\mathbf {\bar B}}_k} {\mathbf P}{{\mathbf {\bar B}}_k}^H \bigr) {{\mathbf C}_k}^{1/2}
\end{math} 
takes the same form as in \eqref{eq:delta_probability}, %but with the zero-forcing directions, 
the  other parameters simplify to 
\begin{math}
\tilde{\mathbf {r}}_k  = 
{{\mathbf C}_k}^{1/2}   {\mathbf b}_k \end{math}
and 
\begin{math}v_k =  {\frac {p_k}{\gamma_k}} - {\sigma}_k^2
\end{math}.
Note that ${\mathbf r}_k =  {\frac {p_k}{\gamma_k}}  \tilde{\mathbf {r}}_k $; this rescaling simplifies the discussion below. 
Using Lemma~\ref{lem:Hassibi}, each chance constraint in \eqref{eq:chance_constraint_for_original_power_load_ZF_b}
can be rewritten in the deterministic form
%The deterministic reformulation of this problem obtained by applying the results of the previous chapter is
%4.3
%\begin{subequations}
%\label{eq:power_load_integral_ZF}
\begin{equation}
%\hspace{-0.5em}\min_{\{p_k\geq0\}}&\quad \Trace \bigl(\mathbf{B}_\text{ZF} \mathbf{P} {\mathbf{B}}^H_\text{ZF} \bigr)\\
%\text{s.t.}&\quad
{\frac{1}{2 \pi}} \int_{-\infty}^{\infty} \frac{e^{\tau_k(\complexunit\omega+\beta)}}{\complexunit\omega+\beta} \frac{e^{-c_k}}{\det({\mathbf I}-(\complexunit\omega+\beta){\mathbf Q}_k)}d\omega
\geq 1 - \epsilon_k, %\quad \quad \forall k, 
\label{eq:chance_constraint_for_integral_power_load_ZF}
\end{equation}
%\end{subequations}
where $\tau_k = -\sigma^2_k$, $c_k = \sum_{m=1}^{M} {\frac{|\tilde a_{km}|^2(i\omega+\beta){\lambda_{mk}}} {1+(i\omega+\beta){\lambda_{mk}}}}$ and ${\mathbf a}_k = - {\mathbf{C}_k}^{-1/2} \hat{\mathbf{h}}_k$.
Here, we let  $(-\mathbf{Q}_k) = \mathbf{V}_k \boldsymbol{\Lambda}_k \mathbf{V}^H_k$ denote
the   eigen decomposition of $(-\mathbf{Q}_k)$, with $\lambda_{mk}$   denoting the 
\revtsev{$m^{th}$} largest eigenvalue, 
%{\color{red} later, $\lambda_{mk}$ are the eigenvalues of $-\mathbf{Q}_k$. We should try to be consistent}
and define 
%eigenvalues  
% arranged in descending order, 
$\mathbf{\tilde{a}}_k = \mathbf{V}^H_k \mathbf{a}_k$ with $\tilde a_{km}$ being the $m^{th}$ element of $\mathbf{\tilde{a}}_k$.
%{\color{red}one sentence on algo}

\subsection{Approximation and residue computation}

Since the $\ell^{th}$ term of the Taylor series expansion of $e^{-c_k}$ has $M$ poles of multiplicity $\ell$, it is complicated to apply residue theory directly to the integral in \eqref{eq:chance_constraint_for_integral_power_load_ZF}. However, if $c_k$ were a constant, then the integrand would take the form of $e^{\tau_k s} G_k(s)$, where $G_k(s)$ is a rational function of $s=\beta+\complexunit\omega$. In that case, the application of residue theory is quite straightforward. 

%A simple way to approximate the integrand so that $c_k$ is constant, is to set the $\mathbf{a}_k=\mathbf{0}$. As a result, $c_k = 0$. Recall from \eqref{eq:proof_reformulation}, that we define $\mathbf{a}_k$ to be such that 
%\begin{math}
%\operatorname{Pr} \bigl(\boldsymbol{\delta}^H_k {\mathbf Q}_k\boldsymbol{\delta}_k+2\operatorname{Re}(\boldsymbol{\delta}^H_k {\mathbf r}_k) + v_k  \geq  0\bigr)
%\end{math}, or in zero-forcing case 
%\begin{math}
%\operatorname{Pr}_{\boldsymbol{\delta}_k} \Bigl(\boldsymbol{\delta}^H_k {\mathbf Q}_k\boldsymbol{\delta}_k+({\tfrac {p_k}{\gamma_k}})2\operatorname{Re}(\boldsymbol{\delta}^H_k \tilde{\mathbf {r}}_k) + v_k  \geq  0\Bigr)
%\end{math},
%is equivalent to 
%\begin{math}
%\operatorname{Pr}\bigl(\|{\boldsymbol{\delta}_k} - \mathbf{a}_k\|_{(-{\mathbf Q}_k)}^2 \leq {\tau}_k\bigr)
%\end{math}.
%It can be seen that if $\mathbf{a}_k = \mathbf{0}$, we can not construct the term that is linear in $\boldsymbol{\delta}_k$; i.e., $({\tfrac {p_k}{\gamma_k}})2\operatorname{Re}(\boldsymbol{\delta}^H_k \tilde{\mathbf {r}}_k)$. In order to make the assumption of $\mathbf{a}_k = \mathbf{0}$ possible, we seek to approximate 
One way to approximate the probability in \eqref{eq:chance_constraint_for_original_power_load_ZF_b} so that $c_k$ is constant is to approximate  
the linear term $2\operatorname{Re}(\boldsymbol{\delta}^H_k \tilde{\mathbf {r}}_k)$ by an appropriate constant value,
which we will denote by $\eta_k$.
That is, we approximate the constraint in \eqref{eq:chance_constraint_for_original_power_load_ZF_b}
 by 
 \begin{equation}
\label{appr_prob_4}
\operatorname{Pr} \Bigl(\boldsymbol{\delta}^H_k {\mathbf Q}_k\boldsymbol{\delta}_k+({\tfrac {p_k}{\gamma_k}})\eta_k + v_k  \geq  0\Bigr) \geq 1-\epsilon_k.
\end{equation}
  Since 
${\boldsymbol{\delta}_k}$ is a zero-mean Gaussian random variable (RV) with identity covariance matrix, the term $2\operatorname{Re}(\boldsymbol{\delta}^H_k \tilde{\mathbf {r}}_k) $ is a zero-mean Gaussian RV with variance $4\|\tilde{\mathbf{r}}_k\|^2 $. 
That immediately suggests choosing $\eta_k$ to be a negative multiple of $2\|\tilde{\mathbf{r}}_k\|$.
As we explain in Appendix~\ref{app:approx}, the choice of the multiple involves a tradeoff between the 
accuracy and conservatism of the constraint. Guided by the experiments in~\cite{Foad_thesis}, we
will choose the multiple to be $-1.3$ in our numerical experiments.
%
%Based on the value of $\|\tilde{\mathbf{r}}_k\|^2$, we can replace $2\operatorname{Re}(\boldsymbol{\delta}^H_k \tilde{\mathbf {r}}_k)$ with an appropriate constant value, $\eta_k$, and then seek to solve the approximated problem; {\color{red} see Section/Appendix}.

%In Section~\ref{sec:const}, we will discuss how to determine a reasonable value for $\eta_k$ and also a possible iterative algorithm. 
%In this section, we assume that
%$2\operatorname{Re}(\boldsymbol{\delta}^H_k \tilde{\mathbf {r}}_k)$ is replaced by a reasonable constant value, $\eta_k$, and propose the algorithm to solve the new approximated problem. 
%With this replacement, 
Given the choice of values for each $\eta_k$, the robust power loading problem in \eqref{eq:power_load_original_ZF_b}
%the constraint in \eqref{power_load_original_ZF_b} is 
can be approximated by
%4.4
 \begin{subequations}
 \label{eq:power_load_approximate_ZF}
\begin{align}
 \hspace{-0.5em}\min_{\{p_k\geq0\}}&\; \Trace \bigl(\mathbf{B}_\text{ZF} \mathbf{P} {\mathbf{B}}^H_\text{ZF} \bigr)\\
%4.4.b
 \text{s.t.}&\;
\operatorname{Pr}_{\boldsymbol{\delta}_k} \bigl(\boldsymbol{\delta}^H_k {\mathbf Q}_k\boldsymbol{\delta}_k + v^{\prime}_k  \geq  0\bigr) \geq 1 - \epsilon_k, %\quad \quad \forall k,  
\label{eq:chance_constraint_for_approximate_power_load_ZF}
\end{align}
\end{subequations}
where
%\begin{math}
%{\mathbf Q}_k  = {{\mathbf C}_k}^{1/2}\bigl( {\frac {p_k}{\gamma_k}} {\mathbf b}_k {\mathbf b}^H_k - {{\mathbf {\bar %B}}_k} {\mathbf P}{{\mathbf {\bar B}}_k}^H \bigr) {{\mathbf C}_k}^{1/2}
%\end{math},
\begin{math}v^{\prime}_k =  {\frac {p_k}{\gamma_k}} \eta_k+ v_k =  {\frac {p_k}{\gamma^{\prime}_k}} - {\sigma}_k^2
\end{math}, and
\begin{math}
\gamma^{\prime}_k = \frac{\gamma_k}{1+\eta_k}
\end{math}.
%The probability on the left hand side of \eqref{eq:chance_constraint_for_approximate_power_load_ZF}  can be written as
%
%
%Now by taking the same steps that were taken in Section~\ref{sec:General}, we can convert the chance constraint in \eqref{eq:power_load_approximate_ZF} to 
%$\operatorname{Pr}(\|{\boldsymbol{\delta}}_k \|^2_{(-{{\mathbf Q}_k})} \leq p_k/\gamma^{\prime}_k - {\sigma}_k^2)$,
%and hence by  using Lemma~\ref{lem:Hassibi} it can be written in the following deterministic form
Using Lemma~\ref{lem:Hassibi}, the deterministic equivalent of the chance constraint in \eqref{eq:chance_constraint_for_approximate_power_load_ZF} is 
%we can convert the approximated probabilistically constrained problem to a deterministic problem with integral constraints, namely 
%4.5
%\begin{subequations}
%\label{eq:power_load_integral_apprixmated_ZF}
\begin{equation}
%\hspace{-0.5em}\min_{\{p_k\geq0\}}&\quad \Trace \bigl(\mathbf{B}_\text{ZF} \mathbf{P} {\mathbf{B}}^H_\text{ZF} %\bigr)\\
%\text{s.t.}&\quad
{\frac{1}{2 \pi}} \int_{-\infty}^{\infty} \frac{e^{\revtsix{v}^\prime_k(\complexunit\omega+\beta)}}{\complexunit\omega+\beta} \frac{1}{\det({\mathbf I}-(\complexunit\omega+\beta){\mathbf Q}_k)}d\omega
\geq 1 - \epsilon_k. %\quad \quad \forall k, 
\label{eq:chance_constraint_for_integral_approximated_power_load_ZF}
\end{equation}
%\end{subequations}
%where %$\tau^\prime_k = 
%${v}^{\prime}_k = {\frac {p_k}{\gamma^{\prime}_k}}-\sigma_k^2$.
%{\color{red}Is $\tau^\prime$ helpful, or just wasteful notation?}
In terms of the goal of developing an efficient algorithm for robust power loading for the downlink, the key
difference between the approximate constraint in \eqref{eq:chance_constraint_for_integral_approximated_power_load_ZF} and the
exact constraint in \eqref{eq:chance_constraint_for_integral_power_load_ZF} is that the
structure of the numerator of the integrand in \eqref{eq:chance_constraint_for_integral_approximated_power_load_ZF} enables an
application of residue theory to simplify the (exact) computation of that  integral. In particular, as shown in
Appendix~\ref{app:residue}, 
\newchanget{when the non-zero eigenvalues of $\mathbf{Q}_k$ are distinct,}
%by applying residue theory to the integral expression of that chance constraint
%we obtain
the integral in \eqref{eq:chance_constraint_for_integral_approximated_power_load_ZF}, and 
hence the probability on the left hand side of \eqref{eq:chance_constraint_for_approximate_power_load_ZF},
is
%4.6
\begin{equation}
\label{eq:fi_hetro}
%\operatorname{Pr}\bigl(\|{\boldsymbol{\delta}}_k \|^2_{-{{\mathbf Q}_k}} \leq p_k/\gamma^{\prime}_k - {\sigma}_k^2\bigr)
%\\ = 
  \begin{cases}
    1 + \sum_{\ell=1}^{\newchanget{K} - 1} {{f}_\ell}_k({\mathbf{P}}) &  \text{if } p_k \geq \gamma^{\prime}_k  {\sigma}_k^2\\
     -{{f}_{r}}_k({\mathbf{P}}) &  \text{if } p_k < \gamma^{\prime}_k  {\sigma}_k^2
  \end{cases} 
\end{equation}  
where %$r=\min(N_t,K-1)+1$,  
%4.7
\begin{multline}
\label{eq:fi_def}
{{f}_\ell}_k({\mathbf{P}})\\ = 
  \begin{cases}
    0 &  \text{if } {{\lambda _{\ell k}}} = 0 \\
     {-\operatorname{exp}\Bigl({({\frac {1}{\gamma^{\prime}_k}} p_k - {\sigma}_k^2){\frac{-1}{{\lambda _{\ell k}}}}}} \Bigr)\frac{1}{\prod_{j \not= \ell} (1 - {{ \lambda _{jk}}}/{{ \lambda _{\ell k}}})} &  \text{otherwise}
  \end{cases}
\end{multline} 
and ${\lambda _{mk}} $ is the $m^{th}$ largest eigenvalue of $(-{\mathbf Q}_k)$.
\newchanget{%Given the structure of ${\mathbf Q}_k$ (cf.~\eqref{eq:delta_probability}), 
As shown in Appendix~\ref{app:residue}, for a large class of 
channel distributions  the non-zero eigenvalues of $\mathbf{Q}_k$ are distinct
with high probability.}
%, the non-zero eigenvalues of which are assumed to be distinct. 
%{\color{red} here is were the $\lambda$s are defined as eigenvalues of $-\mathbf{Q}$. Let's be consistent. Its probably easier to change the earlier ones.}

%%%%%%%%%%%%%%%%%%%%%%%%%%%%%%%%%%%%%%%%%%%%%%%%%
%							Coordinate Descent Algorithm for ZF										%
%%%%%%%%%%%%%%%%%%%%%%%%%%%%%%%%%%%%%%%%%%%%%%%%%
\subsection{Feasible coordinate descent algorithm for \eqref{eq:power_load_approximate_ZF}}
\label{sec:Coordinate_Descent_ZF}
%{\color{red} think about title}
A straightforward approach to exploiting the above analysis is simply to replace integral calculation
that is implicit in Step~5 of Algorithm~\ref{Alg:general_case} by \eqref{eq:fi_hetro}. 
Since \eqref{eq:fi_hetro} can be computed with much less effort than
\eqref{eq:chance_constraint_for_integral_power_load_ZF} this  algorithm incurs
a much lower computational cost
%less costly, in a computational sense, 
than a direct application of Algorithm~\ref{Alg:general_case} to \eqref{eq:power_load_original_ZF_b}. 
\changet{Since the problem in \eqref{eq:power_load_approximate_ZF} can also be viewed in the framework of standard interference
functions~\cite{Yates_interference_functions}, if the initial power allocation is feasible (and we solve for the minimal feasible $p_k$ at each
step), then the algorithm converges to a globally optimal solution to \eqref{eq:power_load_approximate_ZF}.}
The discussion in Appendix~\ref{app:approx} guides the choice of $\eta_k$ so that, with high probability,
solving the approximate problem in \eqref{eq:power_load_approximate_ZF}, with the assistance of \eqref{eq:fi_hetro}, is equivalent to
solving the original problem in \eqref{eq:power_load_original_ZF_b}.

\subsection{Coordinate update algorithm for   approximating  \eqref{eq:power_load_approximate_ZF}}
\label{sec:novel_ZF}
%{\color{red}think about title}

%In this section, we %use the probability of success formula in \eqref{eq:fi_hetro} to 
%develop an alternative approach to finding good solutions to \eqref{eq:power_load_approximate_ZF}. 
%{\color{red} needs to make the argument why this is interesting; direct update of p's, no bisect? also
%might need to state the algorithm formally}
%Unlike our previous algorithms, which start from a feasible point and then seek to reduce each power in turn in such a way that the power allocation remains feasible, this method starts from a power allocation that is not necessarily feasible. Using the iterative algorithm described in this section, the power allocation is iteratively updated in such a way that it often converges to a good solution. 

Although employing the approximation in \eqref{eq:fi_hetro} of the integral in \eqref{eq:chance_constraint_for_integral_power_load_ZF} results in
an algorithm that is significantly cheaper than the generic application of
Algorithm~\ref{Alg:general_case}, each step in the bisection search in Step~5 of
the approximate algorithm requires, among other things, the eigen decomposition of
the current $(-\mathbf{Q}_k)$. In this section we develop an alternate approximate algorithm that
enables direct updating of $p_k$ and enables all the powers to be updated using only one
eigen decomposition of each $(-\mathbf{Q}_k)$. Unlike the algorithms in the previous sections, the iterates
of the algorithm in this section are not necessarily feasible, but the 
power allocation is cyclically updated in such a way that it often converges to a good solution.
For that reason, we will \revtsev{refer} to the algorithm in this section as a coordinate update algorithm, as distinct
from the coordinate descent algorithm in the previous section. 

To develop an approximate cyclic coordinate update algorithm for the problem in 
\eqref{eq:power_load_approximate_ZF}, we begin by observing that ${{f}_\ell}_k({\mathbf{P}})$
in \eqref{eq:fi_hetro} depends on $p_k$ both explicitly, and implicitly through the eigenvalues of
$(-{\mathbf Q}_k)$. To avoid the complexity that this implicit dependence incurs, in the $i^{th}$ cycle
of updates, we will employ the following approximation of ${\mathbf Q}^{(i)}_k$,
\begin{equation}
{\hat{\mathbf Q}}^{(i)}_k = {{\mathbf C}_k}^{1/2}\Bigl( {\tfrac {p^{(i-1)}_k}{\gamma_k}} {\mathbf b}_k {\mathbf b}^H_k - {{\mathbf {\bar B}}_k} {\mathbf P}^{(i-1)} {{\mathbf {\bar B}}_k}^H \Bigr) {{\mathbf C}_k}^{1/2},
\label{eq:Qhat}
\end{equation}
where $ {\mathbf P}^{(i-1)} = \operatorname{Diag}(p^{(i-1)}_1,p^{(i-1)}_2,\dots,p^{(i-1)}_K)$ is
the power allocation at the end of the previous cycle.
%
%
%To begin,    we observe that in the residue expression of the integral in \eqref{eq:fi_hetro}, ${{f}_\ell}_k({\mathbf{P}})$, is a function of $p_k$, through the eigenvalues of $-{\mathbf Q}_k$, and this relationship is difficult to express analytically. That complicates the development of a cyclic coordinate update algorithm. 
%To address that difficulty, at the $k^{th}$ step of the $i^{th}$ cycle we will construct an approximation
%of ${\mathbf Q}^{(i)}_k$ as ${\hat{\mathbf Q}}^{(i)}_k = {{\mathbf C}_k}^{1/2}\bigl( {\frac {p^{(i-1)}_k}{\gamma_k}} {\mathbf b}_k {\mathbf b}^H_k - {{\mathbf {\bar B}}_k} {\mathbf P}^{(i-1)}_k {{\mathbf {\bar B}}_k}^H \bigr) {{\mathbf C}_k}^{1/2}$.  There are two standard variations on how to construct ${\mathbf P}^{(i-1)}_k$, namely,
%%We can construct ${\mathbf P}^{(i-1)}_k$ based on the power allocations in previous cycle for all users; i.e., 
%${\mathbf P}^{(i-1)}_k= {\mathbf P}^{(i-1)} = \operatorname{Diag}(p^{(i-1)}_1,p^{(i-1)}_2,\dots,p^{(i-1)}_K)$,
%and 
%%. The alternative way is to use the updated powers of the current cycle for prior users and previous cycle power allocations for the rest; i.e., 
%${\mathbf P}^{(i-1)}_k = \operatorname{Diag}(p^{(i)}_1,\dots,p^{(i)}_{k-1},p^{(i-1)}_{k},p^{(i-1)}_{k+1},\dots,p^{(i-1)}_K)$. 
%Although the second choice may result in fewer cycles, our numerical results suggest that constructing ${\hat{\mathbf Q}}^{(i)}_k$ based on the first choice of ${\mathbf P}^{(i-1)}_k$ leads to a faster algorithm since   ${\hat{\mathbf Q}}^{(i)}_k$ and its eigenvalues need only be evaluated   once in each cycle. 

With that approximation of $\mathbf{Q}_k^{(i)}$ in place, at the $k^{th}$ step of the $i^{th}$ cycle we 
are looking for a power  $p_k^{(i)}$ that lies  close to the boundary of the feasible set. That is we are
looking for the smallest non-negative $p_k^{(i)}$ such that 
%
%{\color{red}flow needs work}
%According to the strategy in the previous {\color{red}section?}, we are looking for a power allocation near the boundary of feasible set. At the $k^{th}$ step of the $i^{th}$ cycle, the design problem becomes 
%$%\displaystyle{
%\min_{p^{(i)}_k\geq 0} p^{(i)}_k
%%}
%$ subject to 
%4.11
\begin{equation}
\label{eq:novel_condition_power}
  \begin{cases}
    1 + \sum_{\ell=1}^{\newchanget{K}- 1} \hat{{f}_\ell}_k (p^{(i)}_k)  \geq 1 - \epsilon_k &   \text{if } p^{(i)}_k \geq \gamma^{\prime}_k  {\sigma}_k^2\\
   - \hat{f}_{rk} (p^{(i)}_k) \geq 1 - \epsilon_k &   \text{if } p^{(i)}_k < \gamma^{\prime}_k  {\sigma}_k^2
  \end{cases}
 \end{equation}
 where
 %%%4.12
 \begin{multline}
\hat{f}_{\ell k}(p^{(i)}_k)  = \\
%\left\{ 
  \begin{cases}
    0 &  \text{if } {{\hat{\lambda}^{(i)}_{\ell k}}} = 0 \\
     {-\operatorname{exp}\Bigl({({\frac {1}{\gamma^{\prime}_k}} p^{(i)}_k - {\sigma}_k^2){\frac{-1}{{\hat{\lambda}^{(i)}_{\ell k}}}}}} \Bigr)\frac{1}{\prod_{j \not= \ell} (1 - {{ \hat{\lambda}^{(i)}_{jk}}}/{{ \hat{\lambda}^{(i)}_{\ell k}}})} &\text{otherwise}
  \end{cases}
\end{multline} 
and ${\hat{\lambda}^{(i)}_{mk}} $ is the $m^{th}$ largest eigenvalue of $(-{\hat{\mathbf Q}}^{(i)}_k)$. Here we have assumed that the non-zero eigenvalues are distinct.  
 
Since we are looking for the smallest $p^{(i)}_k$, we first consider the second case in \eqref{eq:novel_condition_power} where $p^{(i)}_k < \gamma^{\prime}_k  {\sigma}_k^2$. In this case we look for the smallest $p^{(i)}_k$ that satisfies 
$\hat{f}_{rk}(p^{(i)}_k) \geq 1 - \epsilon_k$, 
%which is equivalent to 
%{\color{red} do we need to relate this to the constraint that $p^{(i)}_k < \gamma^{\prime}_k  {\sigma}_k^2$?}
 %4.13
 \revtsix{namely}
 \begin{equation}
\label{eq:ineq_ZF_2}
%p^{(i)}_k \geq 
\revtsix{\tilde{p}=}
{\gamma^{\prime}_k}{\sigma_k^2}-{\gamma^{\prime}_k}{{\hat{\lambda}}^{(i)} _{r k}}\operatorname{ln}\Bigl((1-{\epsilon_k}){\prod_{j \not= r} (1 - {{{\hat{\lambda}}^{(i)} _{jk}}}/{{{\hat{\lambda}}^{(i)} _{r k}}})} \Bigr).
 \end{equation} 
 \revtsix{If $\tilde{p}\in(0,\gamma^{\prime}_k  {\sigma}_k^2)$, it is
 an admissible solution for \eqref{eq:novel_condition_power}.}
%Since we seek small powers, $p^{(i)}_k$ is chosen such that equality holds in \eqref{eq:ineq_ZF_2}. This $p^{(i)}_k$ is admissible when it lies in the interval of $(0,\gamma^{\prime}_k  {\sigma}_k^2
%)$, which is the assumption that underlies the derivation of \eqref{eq:ineq_ZF_2}. 
%Our numerical experiments show that this condition is not satisfied very often. Therefore, most of the time the other case 
%in \eqref{eq:novel_condition_power} should be considered. 
%{\color{red} that argument needs refinement}

If %the minimum power satisfying \eqref{eq:ineq_ZF_2} 
\revtsix{$\tilde{p}$}
is not admissible, 
%For the other case in \eqref{eq:novel_condition_power}, 
the desired solution is  the smallest non-negative root of $ \sum_{\ell=1}^{r-1} \hat{{f}_\ell}_k(p^{(i)}_k)  + \epsilon_k = 0$
\revtsix{that is not smaller than $\gamma_k^\prime\sigma_k^2$}; cf.~\eqref{eq:novel_condition_power}. Since $ \hat{{f}_\ell}_k(p^{(i)}_k)$ is smooth, any
one of a  number of standard root finding algorithms could be considered. Instead of doing that, we will employ a conservative approximation of the constraint $1 + \sum_{\ell=1}^{r-1} \hat{{f}_\ell}_k(p^{(i)}_k) \geq 1- \epsilon_k$ and show that the resulting problem has a closed-form solution. As $\ell$ increases, the argument of the exponential in \eqref{eq:fi_hetro} becomes more negative,
and hence the magnitude of $\hat{{f}_\ell}_k(p^{(i)}_k)$ decreases. Furthermore, for odd $\ell$, $\hat{{f}_\ell}_k(p^{(i)}_k)< 0$, whereas for even $\ell$, $\hat{{f}_\ell}_k(p^{(i)}_k) > 0$. As a result we have that $\sum_{\ell=2}^{r-1} \hat{{f}_\ell}_k(p^{(i)}_k) \geq 0$.
\revtsix{(Typically, this term will also be}
% and that this term will typically be 
small in comparison to $| \hat{{f}_1}_k(p^{(i)}_k)|$). Therefore, if $p^{(i)}_k$ is chosen such that $1 +  \hat{{f}_1}_k(p^{(i)}_k) \geq 1- \epsilon_k$, then  the outage constraint is guaranteed to hold.
More explicitly, 
\revtsix{if we let}
%$p^{(i)}_k$ is chosen such that 
%{\color{red} do we need to relate this to the constraint that $p^{(i)}_k \geq \gamma^{\prime}_k  {\sigma}_k^2$
%do we need a ``max'' with that involved?}
%4.14
\begin{equation}
\label{eq:ineq_ZF}
 %p^{(i)}_k \geq 
 \revtsix{\breve{p} =}  \gamma_k{\sigma_k^2}-{\gamma_k}{{\hat{\lambda}}^{(i)} _{1k}}\operatorname{ln}\Bigl({\epsilon_k}{\prod_{j \not= 1} (1 - {{{\hat{\lambda}}^{(i)} _{jk}}}/{{{\hat{\lambda}}^{(i)} _{1k}}})} \Bigr),
\end{equation}
\revtsix{then if $\tilde{p}$ is not admissible we choose
$p_k^{(i)}= \max\{\breve{p}, \gamma_k^\prime\sigma_k^2\}$.}
%Since we seek small powers, $p^{(i)}_k$ is chosen such that equality holds in \eqref{eq:ineq_ZF}. 

Having established the iterations above, we need to select the initial powers. 
Using the insight developed in  \cite{Davidson_Foad_ICASSP2013} for a slightly different system,
we will 
%To initialize  the algorithm, we 
chose $p^{(0)}_k$ as if we have equal power allocation with the $k^{th}$ user's parameters. 
\revtsix{That is, to set $p^{(0)}_k$  we let $\mathbf{P}^{(0)}$ take the form $\check{p}\mathbf{I}$,
and determine the value of $\check{p}$ that yields equality in the $k^{th}$ user's approximation
of the outage constraint; cf.~\eqref{eq:chance_constraint_for_approximate_power_load_ZF}. This value can be computed in closed form:}
%{\color{red} this needs more explanation. Perhaps one sentence more}
%For the system considered by \cite{Davidson_Foad_ICASSP2013}, equal power allocation is optimal in the  scenario in which all users have the same 
%$\sigma$, $\gamma$, $\mathbf{C}$ and $\epsilon$. However, for the SINR definition which is used in this thesis, that is not the case. Despite that, we still use the same method for selecting $p^{(0)}_k$. 
%That is, we choose
%4.15
\begin{equation}
\label{eq:initial_power_ZF}
p^{(0)}_k = \frac{\sigma_k^2} {  {1}/{\gamma^{\prime}_k}+{{\tilde{\lambda}}_{1k}} \operatorname{ln}\bigl(\epsilon_k {\prod_{j \not= 1} (1 - {{{\tilde{\lambda}}_{jk}}}/{{{\tilde{\lambda}}_{1k}}}) \bigr)}},
% \quad  \forall k,
\end{equation}
where ${{\tilde{\lambda}} _{mk}} $ is the $m^{th}$ largest eigenvalue of the corresponding $-{\tilde{\mathbf {Q}}}_k = - {{\mathbf C}_k}^{1/2}\bigl( {\frac {1}{\gamma_k}} {\mathbf b}_k {\mathbf b}^H_k - {{\mathbf {\bar B}}_k} {{\mathbf {\bar B}}_k}^H \bigr) {{\mathbf C}_k}^{1/2}$. 
%This choice is based on the insight of \cite{Davidson_Foad_ICASSP2013} for a slightly different system. 
Unlike the coordinate descent 
algorithm in previous chapter and that in Section~\ref{sec:Coordinate_Descent_ZF}, this initial power allocation is not necessarily feasible, but as the coordinates are updated, the power allocation tends to move toward the feasible set. The cyclic updates are terminated once a feasible point is found or if no feasible point is found in a reasonable time. (Feasibility is evaluated using the expression in \eqref{eq:fi_hetro} and comparing with $1-\epsilon_k$.) Our numerical experience suggests that the starting point in \eqref{eq:initial_power_ZF} is particularly effective in that the level of conservatism in the first feasible point tends to be low. In systems where that is not the case, one can use this feasible point to initialize the
algorithm in Section~\ref{sec:Coordinate_Descent_ZF}.
The algorithm developed in this section is summarized in Algorithm~\ref{Alg:coord_update}.
%{\color{red} \bfseries \itshape The algorithm statement needs to be checked, especially since I have changed it
%to a while loop form to match algo 1.}

\begin{algorithm} %[t]
%\caption{Coordinate update algorithm for   approximating \eqref{eq:power_load_approximate_ZF}}
\caption{}
\label{Alg:coord_update}
\begin{algorithmic}[1]
\State Select   $i_{\text{max}}$, and set  each $p_k^{(0)}$ according to \eqref{eq:initial_power_ZF}. 
\revtsev{Set $i=0$.}
	
       %  \For{$i = 1 \to i_{\text{max}}$}
       \While{\revtsev{current power allocation is infeasible and $i\leq i_{\text{max}}$}}
       \State \revtsev{$i=i+1$}
         		\State \revtsev{For all $k$}, construct $\hat{\mathbf{Q}}_k^{(i)}$ according to
		\eqref{eq:Qhat} and compute its eigen decomposition.
      		 \For{$k= 1 \to K$} 
		      %  \State Update $c_k$, $\tau_k$, $\mathbf{Q}_k$ 
       			\State  Compute $\tilde{p}$ %according to equality 
			in \eqref{eq:ineq_ZF_2}.
			\If{$\tilde{p}\in(0,\gamma_k^\prime \sigma_k^2)$} 
			 set $p_k^{(i)} = \tilde{p}$
			\Else $\:$compute \revtsix{$\breve{p}$
			in \eqref{eq:ineq_ZF} and set $p_k^{(i)}= \max\{\breve{p}, \gamma_k^\prime\sigma_k^2\}$}.
			%according to {\color{red} equality?? (do we need the max??)} in \eqref{eq:ineq_ZF}.
			\EndIf
			\EndFor
		\State Evaluate feasibility of current power allocation   using
		\eqref{eq:chance_constraint_for_approximate_power_load_ZF} and \eqref{eq:fi_hetro}.
%		\If{current power allocation is feasible}
%		\State STOP
		\EndWhile
		%\EndFor
\end{algorithmic}
\end{algorithm}

%%%%%%%%%%%%%%%%%%%%%%%%%%%%%%%%%%%%%%%%%%%%%%%%%
%							Discussion on How to Choose $\eta_k$									%
%%%%%%%%%%%%%%%%%%%%%%%%%%%%%%%%%%%%%%%%%%%%%%%%%

%%%%%%%%%%%%%%%%%%%%%%%%%%%%%%%%%%%%%%%%%%%%%%%%%%%%%%%%%%%%%%%%%%%%%%%%%%%%%%%%%%%%%
% 											Performance Evaluation
%%%%%%%%%%%%%%%%%%%%%%%%%%%%%%%%%%%%%%%%%%%%%%%%%%%%%%%%%%%%%%%%%%%%%%%%%%%%%%%%%%%%%
\section{Performance Evaluation}
\label{sec:sims}

In this section, we demonstrate the performance of the proposed algorithms.
For \revtsix{the} general algorithm developed  in Section~\ref{sec:general_algo} (see Algorithm~\ref{Alg:general_case}), 
we will use regularized channel inversion (RCI) beamforming, ${\mathbf{B}}_{\text{RCI}} = {\hat {\mathbf{H}}}^{H} \bigl({\hat {\mathbf{H}}} {\hat{\mathbf {H}}}^H+{\alpha {\mathbf I}_K}\bigr)^{-1}$, and zero-forcing beamforming, ${\mathbf{B}}_{\text{ZF}} = {\hat {\mathbf{H}}}^{H} \bigl({\hat {\mathbf{H}}} {\hat{\mathbf {H}}}^H\bigr)^{-1}$, 
\newchanget{and the (PCSI) beamformers that are obtained by treating the estimated CSI as if it were perfect
and solving \eqref{eq:perfect_CSI_problem},}
as   examples of fixed-direction beamformers. For the RCI case, we specify the regularization parameter, $\alpha$, according to the results of \cite{Swindle} for the full CSI case,
\changet{namely $\alpha=K\sigma^2$ where $\sigma^2$ is the noise variance at each 
receiver (which will be assumed to be the same)}. %In this case, $\alpha=0.03$.
%The value of $\alpha$ is given in Table~\ref{tabel:description}.
%
%{\color{red} for systems based on Algo 1, how many times was the iterative doubling trick tried? Foad will check}
%
%
 For the algorithms proposed in Section~\ref{sec:ZF}, which seek good solutions to the approximated problem in \eqref{eq:power_load_approximate_ZF} for the \revtsix{nominally} zero-forcing case
 (see Section~\ref{sec:Coordinate_Descent_ZF} and Algorithm~\ref{Alg:coord_update}),
 based on analysis in \cite{Foad_thesis} the value of the parameter $\eta_k$ is chosen to be
 $-1.3(2\|\tilde{\mathbf{r}}_k\|)$.
%  we provide four different approaches. Two of them implement the coordinate descent algorithm proposed in Section~\ref{sec:Coordinate_Descent_ZF}, one having a pre-specified value of $\eta_k$ while the other iteratively changes the value of $\eta_k$ as described in Section~\ref{sec:const}. The other two demonstrate the algorithm for power allocation based on the essence of the probability of success described in Section~\ref{sec:novel_ZF}. Again, one has a specific value for $\eta_k$ while the other iteratively changes that value. In the schemes in which $\eta_k$ should be pre-specified, we choose a reasonable value according to the analysis in Appendix~\ref{App:eta}. The values are given in Table~\ref{tabel:description}.

%In order to 
\revtsix{We will} compare our algorithms to  \revtsix{several}  existing methods, 
%we consider 
\revtsix{including} the chance-constrained robust precoding method of \cite{Ken_outage}, which is  based on  rank-one relaxation and convex restriction  (RAR) and was formulated in \eqref{eq:cons_precoder_RARstyle}. This method is a full precoder design method,
and hence the beamformers and the power allocation are designed jointly.
\revtsix{In terms of power allocation methods, we will}
% We also 
compare against the adaptation of the RAR approach to  robust power loading  of   that was formulated in \eqref{eq:cons_powerload_RARstyle},
\revtsix{and the MSE-based fixed-point iteration method in \cite{Vucic_Boche}.}
 %For solving these two convex problems, we use the conic optimization solver SeDuMi, implemented through the now popularized and convenient parser software CVX \cite{grantcvx}. 
%{\color{red} use proper reference}
For convenience, a summary of methods that will be considered, and the labels that will describe them, is provided in Table~\ref{tabel:description}.
%
%{\color{red}why do we only compare to RAR? need to make the case for that. why not other
%methods? Vucic and Boche SPL paper? Foad is thinking about this.}

In the following simulations, we consider an environment with $N_t=3$ transmit antennas, $K=3$ users, i.i.d.\ Rayleigh fading channels, and the receivers' noise sources are modeled as zero-mean, additive, white, and Gaussian with variance $\sigma^2_k = \sigma^2 = 0.01$. 
In our simulations the errors in the BS's estimates of the channels are generated by explicitly performing
uplink training with orthogonal training sequences 
\newchanget{and linear MMSE channel estimation~\cite{Biguesh_Gershman_training}.}%
%\newchangem{In order to evaluate the proposed approach under some model mismatch, we will
%employ linear MMSE channel estimation~\cite{Biguesh_Gershman_training}.}%   
%\newchangem{and the least-squares estimator, which is the minimum variance estimator in this setting~\cite{Biguesh_Gershman_training}}
%\footnote{\newchangem{\bfseries If we make this change, we will have to re-do all the simulations! I know that that may be a lot of work, but the programming effort should not be very much. Just changing the MMSE estimator to the LS estimator should require a simple removal of a regularization term based on the noise variance. I can help on that front if you can give me the line of code that you use to determine the MMSE channel estimate}}
%\newchanget{and the linear MMSE estimator (which is
%unbiased in this setting).}%
\footnote{\revtsix{The TDD ``ping-pong'' time is assumed to be short enough and the RF calibration
good enough (cf.~\cite{kaltenberger2010relative}) for the   base-band equivalent channels to be reciprocal.}}
%{\color{red} mention TDD here}
\newchangem{Even though the linear MMSE estimator 
(of a given realization of the channel) is only asymptotically unbiased (as the training SNR increases),}
for  the purposes of   robust \newchangem{power loading} the
%\newchangem{t
%this means that the 
%\newchangem{In this setting, the} 
uncertainties in the channel estimates 
%According to the results of Section~\ref{sec:Sys_TDD} for TDD systems in the case of i.i.d.\ Rayleigh fading channels,
%he uncertainty in the channel estimation 
\newchangem{will} be modeled by   Gaussian random vectors  with zero mean.
\newchangem{The error covariance for the linear MMSE estimator in this setting is} 
% and covariance 
${\mathbf C}_k=\mathbf{C} = \sigma^2_e{\mathbf I}$,
where 
$\sigma_e^2= %\newchangem{\frac{\sigma_{\text{BS}}^2}{ L_{\text{UT}}P_{\text{UT}}}}$
 \frac{\sigma_{\text{BS}}^2}{\sigma_{\text{BS}}^2 + L_{\text{UT}}P_{\text{UT}}}$,
with $\sigma_{\text{BS}}^2$ being the noise variance at the BS's receiver for the uplink,
$L_{\text{UT}}$ being the length of the training sequence, and $P_{\text{UT}}$ being the power of the
training sequence, \newchangem{and this covariance will be used in the model of the error}.
%; e.g.,~\cite{Foad_thesis}.
In terms of performance specifications, the probability of outage is set to be ${\epsilon_k}=\epsilon=0.05$ for all users and a universal SINR target is defined; i.e., $\gamma_k = \gamma$.

%%%%%%%%% Table 5.1
\begin{table}
\centering
    \caption{Description of methods}
     \begin{tabular}{ll}
       \toprule
    \textbf{Method} & \textbf{Description} \\ 
    %\hline \hline
    \midrule
    \newchanget{PCSI-General} & 
    \newchanget{Robust power loading (RPL) based on the original} \\
    & \newchanget{formulation for the ``perfect CSI'' directions.}\\
    &\newchanget{Solved using Algorithm~\ref{Alg:general_case}.} \\
    \midrule    
    RCI-General
    %-Method ($\alpha=0.03)$ 
    & \newchanget{RPL} based on the original formulation for the \\
   &  RCI directions 
     with  $\alpha = \changet{K\sigma^2}$. % 0.03$.   
     \\ &  
      Solved using Algorithm~\ref{Alg:general_case}. % that seeks to solve the problem in \eqref{eq:power_load_integral} according to Algorithm~\ref{Alg:general_case} in Chapter~\ref{chap3:General}.}
    \\ %\hline \hline
    \midrule 
%2 
ZF-General & RPL based on the original formulation for the \\ & ZF directions. 
   Solved using Algorithm~\ref{Alg:general_case}.
% that seeks to solve the problem in \eqref{eq:power_load_integral} according to Algorithm~\ref{Alg:general_case} in Chapter~\ref{chap3:General}.} \\ \hline \hline
\\ \midrule
%3
ZF-CoordDescent & RPL  for the ZF directions
based on the algorithm \\
& in Section~\ref{sec:Coordinate_Descent_ZF}, which seeks good solutions to
the \\
 & approximated problem in \eqref{eq:power_load_approximate_ZF} with \\ &  $\eta_k=-1.3(2\|\tilde{\mathbf{r}}_k\|)$.
 %that seeks to solve the approximated problem in \eqref{eq:power_load_approximate_ZF} according to Algorithm~\ref{Alg:Descent_ZF} in Section~\ref{sec:Coordinate_Descent_ZF}, where $\eta_k = - \eta^{\prime}\times2\|\tilde{\mathbf{r}}_k\|$ and $\eta^{\prime}=1.3$.} \\ \hline \hline
 \\ \midrule
 %4
% \normalsize{ZF-App-CordDescent-Iterative($\eta_k$)} & \normalsize{RPL algorithm in the ZF direction that seeks to solve the approximated problem in \eqref{eq:power_load_approximate_ZF} according to Algorithm~\ref{Alg:Descent_ZF} in Section~\ref{sec:Coordinate_Descent_ZF}, where $\eta_k$ iteratively changes as described in Appendix~\ref{App:zeta}.} \\ \hline \hline
%5
ZF-CoordUpdate & RPL for the ZF directions based on Algorithm~\ref{Alg:coord_update},\\
&
which seeks good solutions 
  to the approximated \\ &  problem in \eqref{eq:power_load_approximate_ZF} with $\eta_k=-1.3(2\|\tilde{\mathbf{r}}_k\|)$. 
%according to the method proposed in Section~\ref{sec:novel_ZF}, where $\eta_k = - \eta^{\prime}\times2\|\tilde{\mathbf{r}}_k\|$ and $\eta^{\prime}=1.3$.} \\ \hline \hline
 %6
 %\normalsize{ZF-App-CordUpdate-Iterative($\eta_k$)} & \normalsize{RPL algorithm in the ZF direction that seeks to solve the approximated problem in \eqref{eq:power_load_approximate_ZF} according to the method proposed in Section~\ref{sec:novel_ZF}, where $\eta_k$ iteratively changes as described in Appendix~\ref{App:zeta}.} \\ \hline \hline
 %7
\\ \midrule
RAR & Robust precoding based on the rank  relaxation   
\\ & and convex restriction (RAR) method of \cite{Ken_outage}. \\ &The resulting SDP  was formulated in \eqref{eq:cons_precoder_RARstyle}. \\ \midrule 
  %8
  \revtsev{ZF-SDP} & RPL for the ZF directions based on the adaptation 
  \\ &  of the RAR method of \cite{Ken_outage}   to the  power loading  \\ & problem that was formulated in \eqref{eq:cons_powerload_RARstyle}.\\ 
  \midrule
  \revtsix{Fixed-point} & RPL for the ZF directions based on the fixed-point 
  \\ & approach in \cite{Vucic_Boche}.\\
%  \midrule
  %{\color{red} adaptive $\eta_k$?} & {\color{red} actually, I don't think so. Too complicated to explain; computational cost of implementing it}
    \bottomrule
    \end{tabular}
     \label{tabel:description}
\end{table}
\subsection{Performance Comparisons Against SINR Requirements}
\label{sec:SINR}

  %%%%%%%%%% Fig 5-2

\begin{figure} 
\centering
\includegraphics[width=\figwidth]{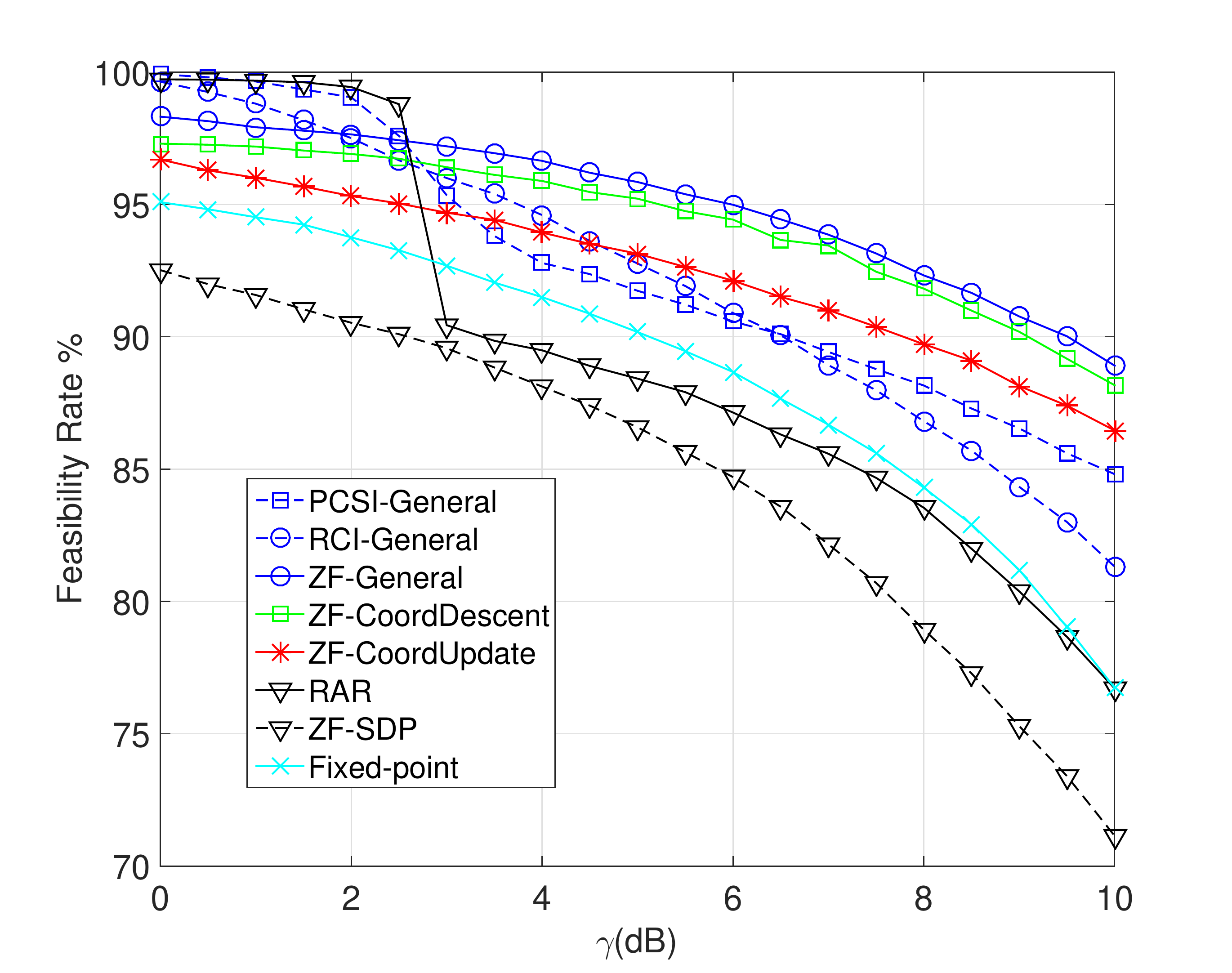}
\caption{\changet{Percentage of successful designs} for different methods in the environment where $N_t=K=3$, ${\mathbf C} =0.002{\mathbf I}$, $\epsilon = 0.05$, and $\sigma^2=0.01$.}
% {\color{red} change y-axis label to ``Success ratio''}}
%{\color{red} don't put the $\alpha$ or $\eta^\prime$ values in the legend. They
%are in the table. This goes for all figures.
%Also, on all legends change ``Vucic \& Boche'' to ``Fixed-point''}}  
\label{fig:Feas_SINR_Fixed}
\end{figure}

\begin{figure}
\centering
\includegraphics[width=\figwidth]{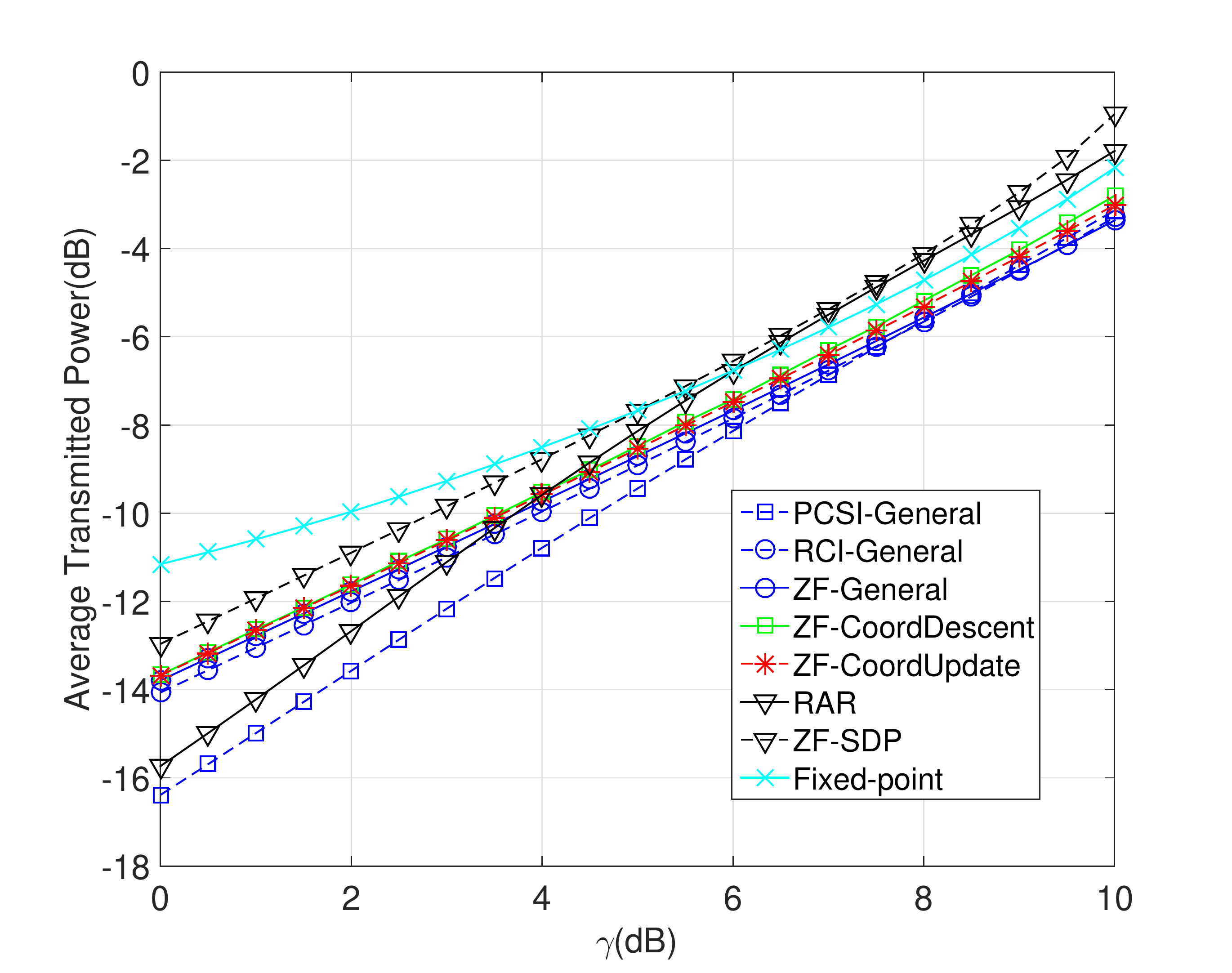}
\caption{Power transmission performance for different methods in the environment where $N_t=K=3$, ${\mathbf C} =0.002{\mathbf I}$, $\epsilon = 0.05$, and $\sigma^2=0.01$;
\revtsix{i.e., the receiver noise power is $-20$dB.}}  \label{fig:Trans_SINR_Fixed}
\end{figure}

In this section, we randomly generated 10,000 realizations of the set of i.i.d.\ Rayleigh fading channels $\{{{\mathbf{h}}}^H_k\}^{K}_{k=1}$. In the uplink training phase, we assumed that there were no peak power constraints on the uplink and we set $L_{\text{UT}} = 1$ and $P_{\text{UT}} = 4.99$ so that with the noise variance at the base station being $\sigma^2_{\text{BS}} = 0.01$, the variance of the channel estimate is $\sigma^2_e = 0.002$. (The choice of $L_{\text{UT}}=1$ is consistent with the results of \cite{hassibi2003much}.) We examined the performance of each \changet{design} method as the SINR requirement of the users, $\gamma$, increases from 0~dB to \changet{10}~dB. 
%{\color{red} extend to 10 dB}
For each set of channel estimates and for each value of $\gamma$, we determined whether each 
\changet{method 
yields a successful design, in the sense that it} generates a  
%
%(or a precoder)
\changet{precoder that satisfies the outage constraints in the original formulation in \eqref{eq:original_problem}. (For the power loading algorithms, the formulation in \eqref{eq:power_load_original} is equivalent.)
%
% (or \eqref{eq:original_problem}, respectively).
Satisfaction of the constraints was determined by evaluating the outage probability using Lemma~\ref{lem:Hassibi}; cf. \eqref{eq:chance_constraint_for_integral_power_load}. In other words, to be deemed ``successful'' for a given set of channel estimates, a design method must
produce a solution that is not only feasible for the optimization problem on which the design is based, but is also feasible
for the original robust precoding problem in \eqref{eq:original_problem}.}%
\footnote{\newchanget{More specifically, when the proposed General algorithm produces a feasible solution to 
\eqref{eq:power_load_integral}, that solution
is guaranteed to be feasible for \eqref{eq:original_problem}. Similarly, when the ZF-SDP method
produces a feasible solution to \eqref{eq:cons_powerload_RARstyle}, 
and when the Fixed-point algorithm produces a solution, those solutions are also guaranteed to be feasible for \eqref{eq:original_problem}.
For the RAR precoding method, a feasible solution to \eqref{eq:cons_precoder_RARstyle} in which all the matrices $\mathbf{U}_k$ have rank one
generates a feasible solution to  \eqref{eq:original_problem}. (If not all of those matrices have rank one, then
one can use a variety of techniques to try to generate a feasible solution to  \eqref{eq:original_problem} 
from the solution to \eqref{eq:cons_precoder_RARstyle}.) For the ZF-CoordDescent and ZF-CoordUpdate methods, the use of the
approximation in \eqref{appr_prob_4} to obtain significant reductions in the computational cost means that when these algorithms produce
feasible solutions to \eqref{eq:power_load_approximate_ZF}, one has to test whether that solution is feasible for  \eqref{eq:original_problem}.}}
%Similarly, the ``success'' of solutions generated  by the Fixed-point algorithm must be checked by determining whether
%those solutions are feasible for \eqref{eq:original_problem}.}}
%
%
%Feasible solutions to ***
% \changet{(The robust precoders generated by the RAR method were evaluated analogously using \eqref{eq:original_problem}.)}
%that guarantees that the probabilistic SINR constraints are satisfied. 
%\changet{Since these designs}
In Fig.~\ref{fig:Feas_SINR_Fixed}, \revtsev{for each design method} 
we plot the percentage of channel realizations for which %each design generated 
\changet{a successful design was obtained.}%
\footnote{\newchangem{Since the training SNR on the uplink is reasonably high, the corresponding results for least squares channel estimation, which is the  minimum variance unbiased estimator in this setting, are indistinguishable from the results 
%for the case of 
%linear MMSE estimation 
in Fig.~\ref{fig:Feas_SINR_Fixed} at the scale of that figure.}}
%a feasible solution \revtsev{was obtained}.} 
%Then, we subsequently selected all the realizations from the set of 10,000 for which all methods provide a feasible solution for all simulated SINR targets.
In Fig.~\ref{fig:Trans_SINR_Fixed}, \revtsev{we plot, against $\gamma$, the average transmission power over the \changet{7,107}
%8,212 
channel set realizations for which all methods provided a \changet{successful design} at all the considered SINR targets}.

From Fig.~\ref{fig:Feas_SINR_Fixed}, it can be seen that by tackling the power loading problem directly (or closely), %even with suboptimal algorithms based on only a single starting point, 
%{\color{red} is it true that there was only one starting point? Foad said no. Will need to adjust accordingly}
 the proposed power loading methods are able to satisfy the QoS constraints more often than the existing power loading method \revtsix{that} is based on an optimal solution  to a tractable conservative approximation  of the problem \changet{(ZF-SDP)}; \revtsix{cf. \eqref{eq:cons_powerload_RARstyle}}.
% {\color{red} need to extend figures to 10dB}
 %and more often than the fixed-point method in \cite{Vucic_Boche}.} 
 This is because the approximation made in that method  \revtsix{can be} quite conservative.%
 \footnote{\changet{The approximation made in the ZF-SDP method involves approximating the outage constraint by the constraint that the design guarantee zero outage for all uncertainties up to a specified size,
 where that size is chosen so that the probability of larger uncertainties is less than the specified outage probability.}}
 \changet{That conservatism can also be seen in Fig.~\ref{fig:Trans_SINR_Fixed}: For  the
 channel realizations for which all methods produce a successful design, the ZF-SDP method uses a
 larger transmission power than the proposed methods.}
 \changet{Figs~\ref{fig:Feas_SINR_Fixed} and~\ref{fig:Trans_SINR_Fixed}
 also show that the} proposed methods  satisfy the QoS constraints more often than the fixed-point method
 in \cite{Vucic_Boche}, \changet{and that they expend less power in doing so.}
 
 \newchanget{Among the ``General'' methods, it is interesting to observe the role that the choice of beamforming directions has on the performance. When the SINR targets are low, the ``perfect CSI'' directions provide the best performance of all the methods we have considered, but when the SINR targets are higher, the nominally ZF directions provide
 the best performance. Furthermore, at higher SINR targets the combinations of the ZF directions and the
 computationally more efficient approximate power loading algorithms provide better performance than the
combinations of the PCSI or RCI directions and the General power loading algorithm.}
  
 What is perhaps more interesting is that for higher SINR targets, the proposed power loading methods provide better performance than the \changet{RAR} robust precoding method   \cite{Ken_outage}, which is formulated in \eqref{eq:cons_precoder_RARstyle}, despite the fact that \changet{the RAR} method has many more degrees of design freedom.%
 \footnote{The RAR method designs the beamformers and power allocation jointly, \newchanget{whereas}
  in the proposed \newchanget{methods, the ZF-SDP method, and the Fixed-point method,   the
 beamformers are fixed.}}
 %methods in which 
 %the beamformers are fixed.}
 %\changet{If the RAR method generates a solution in which each $\mathbf{U}_k$ has rank one, then the design is a success
% in the sense that the corresponding power-loaded beamformers are guaranteed to satisfy the
% outage constraint in the original problem in \eqref{eq:original_problem}.}}
  Once again, this is due to the fact that the techniques used \changet{in the RAR method}
  %in the proposed method for zero-forcing directions 
  to convert the chance constraints \changet{in}to deterministic constraints 
  \changet{can be quite conservative}.%  %are much less conservative.%
 \footnote{\changet{Like the ZF-SDP method, the RAR method is based on a ``zero-outage region''
 approximation of the outage constraint. For lower SINR targets the extra degrees of design freedom in the RAR
 method overcome the conservatism in this approximation, but for higher SINR targets 
 %it appears that the conservatism
 %of this approximation 
 %of the outage constraint 
the extra degrees of design freedom only provide small performance gains over the ZF-SDP method.}}
 \revtten{(An explicit example of the reduced  conservatism of the proposed methods is available in \cite{Foad_thesis}.)}
% {\color{red} we might need more discussion here, especially regarding the drop in performance of the RAR method}
 %{\color{red} mention foad thesis for histograms}
 %\changet{The conservatism of the RAR method also appears to be the cause of the rapid degradation of the RAR method at SINR targets
% between 2.5 and 3~dB. Above that threshold the percentage of channel realizations  for which the SDP in \eqref{eq:cons_precoder_RARstyle} is infeasible is
% much larger than below it. The results for the other methods indicate that for many of these realizations there are precoders that are
 %feasible for the original problem in \eqref{eq:original_problem}. The conservative approximations used to obtain the RAR method 
% appear to exclude these precoders.}   
 
 \changet{We have performed analogous experiments  systems with more antennas and users (4, 5 and 6), and the general structure of Figs~\ref{fig:Feas_SINR_Fixed} and \ref{fig:Trans_SINR_Fixed} and the observations made in the previous two paragraphs apply in those cases, too.}
 % The threshold at which the RAR method breaks down is higher, but a similar thresholding effect occurs.} 

 %%%%%%%%%% Fig 5-3

 %%%%%%%%%%%%%%%%%%%%%%%%%%%%%%%%%%%%%%%%%%%%%%%%%%%%%%%%%%%%%%%%%%%%%%%%%%%%%%%%%%%%%
% 						Performance Comparisons Against Uncertainty Size
%%%%%%%%%%%%%%%%%%%%%%%%%%%%%%%%%%%%%%%%%%%%%%%%%%%%%%%%%%%%%%%%%%%%%%%%%%%%%%%%%%%%%

\subsection{Performance Comparisons Against Uncertainty Size}
\label{sec:Uncer}

\begin{figure}
\centering
\includegraphics[width=\figwidth]{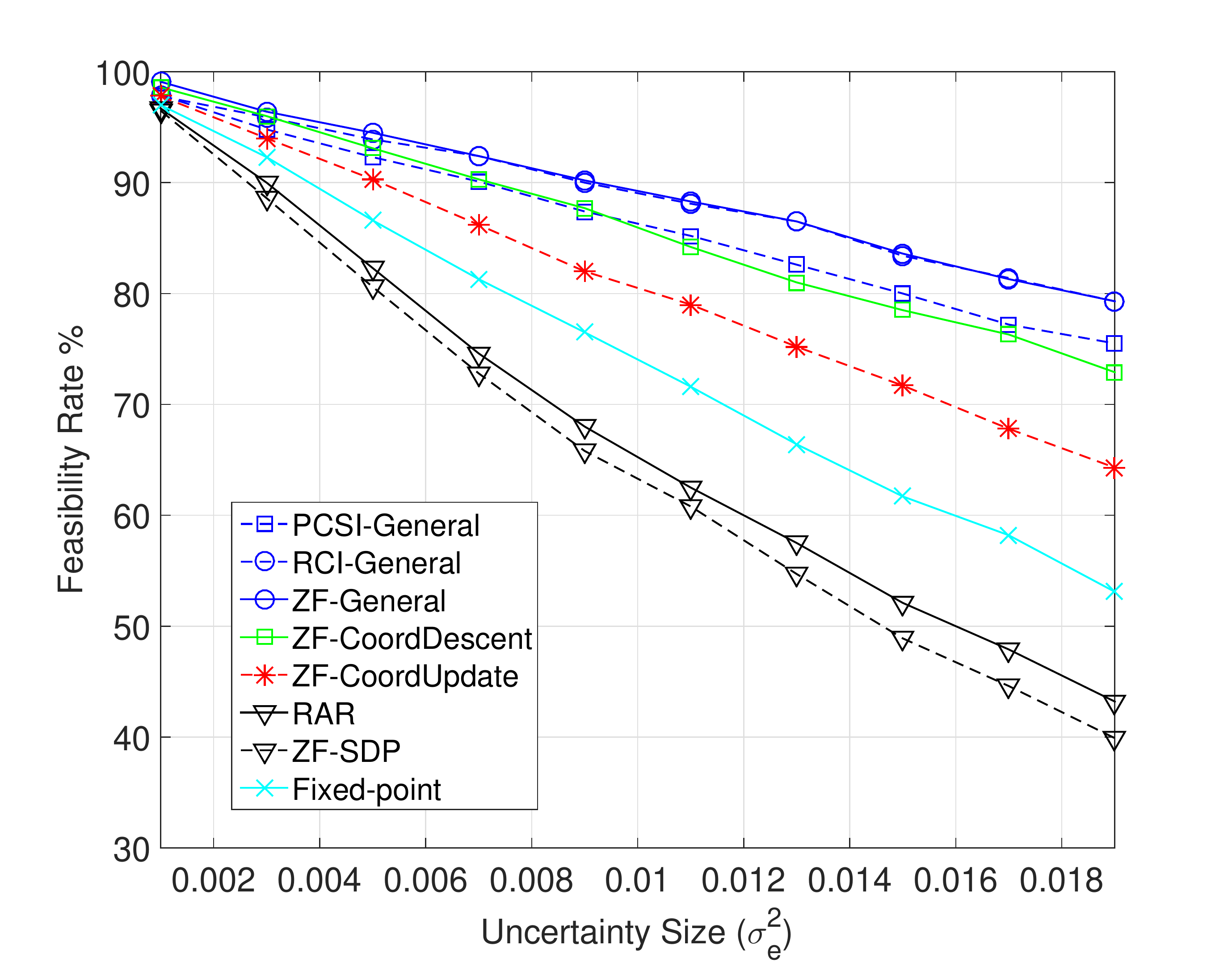}
\caption{\changet{Percentage of successful designs} for different methods in the environment where $N_t=K=3$, $\gamma = \text{3~dB}$, $\epsilon = 0.05$, and $\sigma^2=0.01$. %{\color{red} move x label down in Matlab}
}  \label{fig:Feas_Uncer_Fixed}
\end{figure}

\begin{figure} 
\centering
\includegraphics[width=\figwidth]{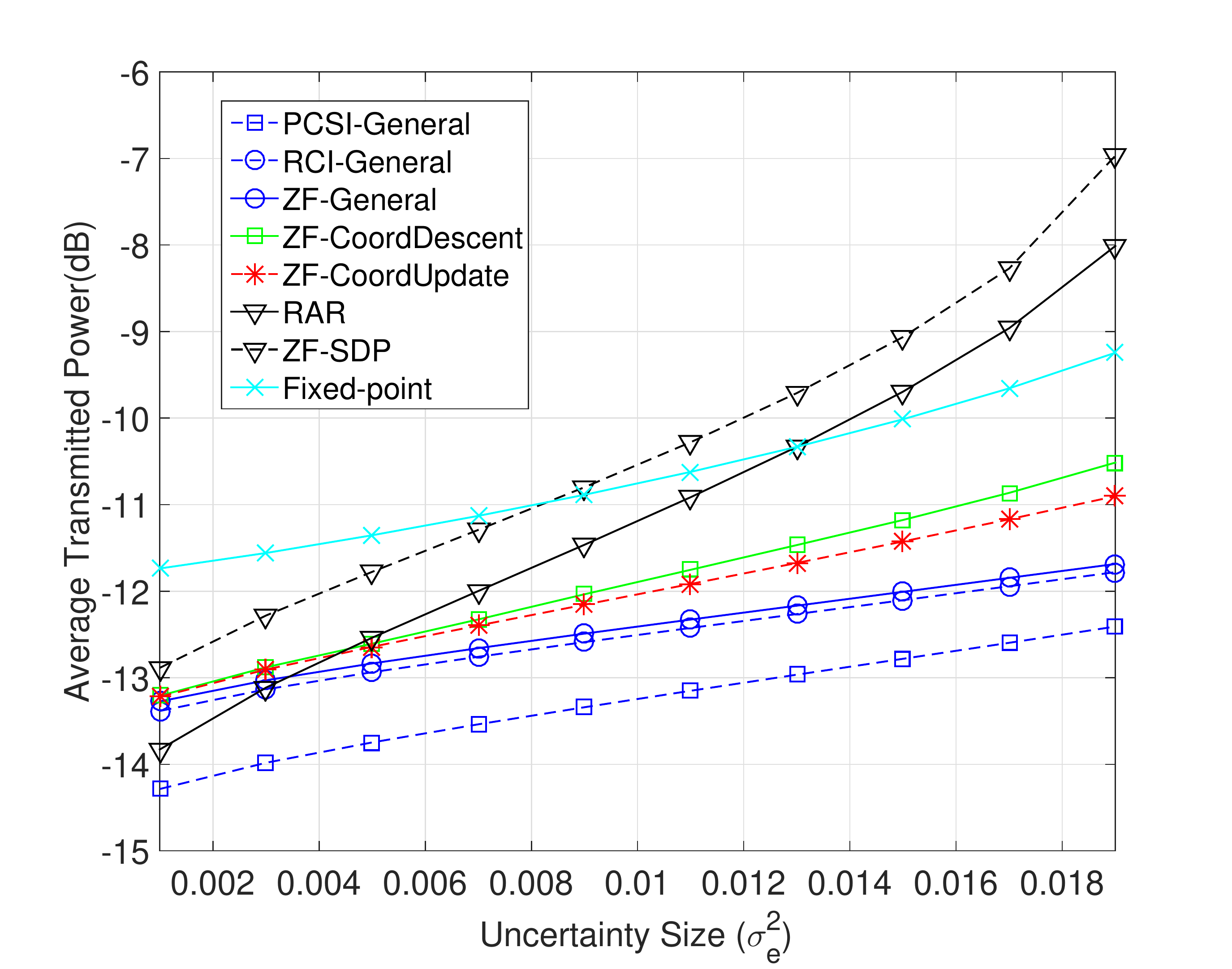}
\caption{Power transmission performance for different methods in the environment where $N_t=K=3$, $\gamma = \text{3~dB}$, $\epsilon = 0.05$, and $\sigma^2=0.01$; 
\revtsix{i.e., the receiver noise
power is $-20$dB}. %{\color{red} move x label down in Matlab}
}  \label{fig:Trans_Uncer_Fixed}
\end{figure}

In the experiments in this section, we randomly generated 1,000 realizations of the set of i.i.d.\ Rayleigh fading channels $\{{{\mathbf{h}}}^H_k\}^{K}_{k=1}$, and similar to the previous simulation, we obtained the CSI at the BS through uplink training. 
In the uplink we choose $L_{\text{UT}}=1$, and 
%Analogous to the experiment in Section~\ref{sec:SINR}, we chose $L_1 = 1$ and 
chose $P_{\text{UT}}$ 
to vary the %
%such that required 
variance of the channel uncertainty, $\sigma^2_e$.
% is produced according to \eqref{eq:MMSE_VAR_BS}. 
In this experiment we set the SINR target of the users to 3~dB. 
In Fig.~\ref{fig:Feas_Uncer_Fixed} we plot the percentage of channel set realizations for which
each algorithm 
%is able to generate a solution that satisfies the constraints, 
\changet{generates a successful design,}
as $P_{\text{UT}}$ is
decreased (and hence $\sigma_e^2$ increases).
In Fig.~\ref{fig:Trans_Uncer_Fixed} we plot the average power that each algorithm requires 
over the 400 channel set realizations for which all methods provide a \changet{successful design} for all the considered
values of $\sigma_e^2$.
\changet{Figs~\ref{fig:Feas_Uncer_Fixed} and \ref{fig:Trans_Uncer_Fixed} show that
%in comparison to the existing methods, 
as the variance of the channel uncertainty increases, 
the performance of the proposed methods degrades much more slowly than that
of the existing methods.}
% significantly less sensitive to increases in the
%variance of the uncertainties.}

%Then, we seek to plot feasibility rate and average transmission power as described in previous section; see Fig.~\ref{fig:Feas_Uncer_Fixed} and Fig.~\ref{fig:Trans_Uncer_Fixed}, respectively. Note that we plot the average transmitted power for $400$ sets of channels for which all methods provide feasible solution.

  %%%%%%%%%% Fig 5-4

 %%%%%%%%%% Fig 5-5

\subsection{\changet{Computational Cost Comparisons}}

As illustrated in Figs~\ref{fig:Feas_SINR_Fixed}--\ref{fig:Trans_Uncer_Fixed}, among the proposed algorithms, the %RCI-General and ZF-General
\newchanget{``General''}
algorithms provide the best performance. This is to be expected because they tackle the original problem without
any approximation. However, these algorithms are also the most computationally expensive
\changet{of the proposed algorithms}. In   \changet{the $i^{th}$} cycle 
they require the evaluation of $\sum_k N^{\changet{(i)}}_{\text{bisect},k}$ integrals of the form in \eqref{eq:bisect_interval},
where $N^{\changet{(i)}}_{\text{bisect},k}$ is the number of bisection steps needed to update $p_k$
\changet{in the $i^{th}$ cycle}. 
\changet{Each of those integrals involves the computation of an eigen decomposition of a  Hermitian symmetric matrix of
size $N_t\times N_t$, which requires $O(N_t^3)$ operations, and the evaluation of a scalar indefinite integral. 
To evaluate the total computational cost of this algorithm
%or any of the other algorithms we have considered, 
we also need to 
 analyze 
 %the number of iterations required, 
 $N^{\changet{(i)}}_{\text{bisect},k}$ and the number of coordinate descent cycles. This is a substantially more difficult task, 
but we can say that in our numerical experiments for   SINR targets of 5~dB, the 
median  of the total number of integrals to be computed,
 $\sum_i\sum_k N^{(i)}_{\text{bisect},k}$,   was 21 for the RCI beamformers and 22 for the ZF beamformers.
 %(The corresponding averages were 20.3 and 20.7, respectively.)
 %{\color{red}was, on average, XX, where the average was computed over ????}.
 %was {\color{red}typically??} around {\color{red} XX. this needs some more explanation}
Furthermore, in these
%but we can say that in our 
\textsc{Matlab}-based numerical experiments, the run-times of our General algorithms were}
%In our experience, the cost of these algorithms is 
of similar magnitude to \changet{those} of the RAR and ZF-SDP methods, 
\changet{although our implementations of the RAR and ZF-SDP methods used compiled code to solve the SDPs. As a more analytical comparison,
we observe that each interior point iteration in the SDP solver requires $O(N_t^6)$ operations.}
%which require the solution of a single SDP to obtain the transmitter parameters.

The ZF-CoordDescent algorithm is significantly faster \changet{than the General 
%and SDP-based 
methods}, with
%\revtnine{an eigen decomposition of a matrix of size $N_t\times N_t$ replacing each integral}.
  the key computational task in each cycle being
 the computation of \revtnine{$\sum_k N^{\changet{(i)}}_{\text{bisect},k}$} eigen decompositions,
 \changet{each of which requires $O(N_t^3)$ operations.}
 % of matrices of size $N_t\times N_t$.
%\revtnine{rather than that many integrals}. 
%\changet{Given the approximations involved, the value for $N^{\changet{(i)}}_{\text{bisect},k}$ 
%for this algorithm may be different from
%the corresponding value in the ``General'' algorithm, and in 
\changet{In our numerical experiments 
for  SINR targets of 5~dB, the median of
the total number of bisection steps, $\sum_i\sum_k N^{\changet{(i)}}_{\text{bisect},k}$,
was~59. Although that number is larger than in the General case, the cost per bisection step is smaller because there is no integral to calculate.} 
%{\color{red} averaged over ??? was  {\color{red}xx}.)}
The cost-per-cycle of
the ZF-CoordUpdate algorithm is \revtnine{significantly} smaller \revtnine{than} 
\changet{that of the ZF-CoordDescent algorithm}, with the key task in each cycle being the computation of \revtnine{only} $K$   eigen decompositions, \changet{each of which requires $O(N_t^3)$ operations.}
\changet{In our numerical experiments for SINR targets of 5~dB, the median number of
cycles of the ZF-CoordUpdate algorithm 
was only~2.}
%terminated in an average of
%xx cycles, and hence an average of Kxx eigen decompositions were required.}
%\changet{Furthermore, our numerical experience suggests that this algorithm requires far fewer
%{\color{red} think about this what should I say here.}}
% \revtnine{of matrices of size $N_t\times N_t$}. 
In  \changet{those \textsc{Matlab}-based} experiments, the run-times of the 
\changet{ZF-CoordDescent and ZF-CoordUpdate} methods \changet{were} about an order of magnitude
lower than that of the \newchanget{``General''}, RAR and ZF-SDP methods. 

Of the methods considered, 
%\changet{Even though its computational cost per cycle is also $O(N_t^)$, 
the fixed-point method was the fastest in our experiments.
%, as each iteration
%can be performed without matrix operations.
%{\color{red} we will need to be more specific here}
\changet{Even though its computational cost per cycle is also $O(N_t^3)$, 
and the median number of cycles for SINR targets of 5~dB was~19, in practice it}
%{\color{red} and its average number of cycles was ???}, in practice it} 
was significantly faster than the proposed ZF-CoordUpdate algorithm.
\changet{That said,} proposed algorithm does provide somewhat better performance,
\changett{and expends less power in doing so.}

\section{Conclusion}
\label{sec:conc}

\revteight{In this paper we have developed algorithms for robust power loading in the MISO downlink
beamforming scenario in the presence of Gaussian uncertainties in the base station's estimates of the 
channels. These algorithms are well-suited to systems that operate in a TDD manner. 
In contrast to many of the existing approaches, the proposed algorithms are based on a precise 
deterministic characterization
of the probability of outage that is induced by the uncertainty, rather than conservative deterministic
characterizations. The precise characterization was incorporated into a cyclic coordinate descent
algorithm that can be viewed within the framework of standard interference functions and hence
is guaranteed to converge to a globally optimal solution. 
Insight into the deterministic characterization then led to the development of approximations that
yield much faster algorithms that retain much of the performance of the original algorithm.
The performance of the proposed algorithms was examined in a number of simulation studies, and
they were shown to have highly desirable performance characteristics, especially in cases of higher SINR
targets and larger uncertainties. In particular, the proposed algorithms provided significantly better performance
than those based on conservative approximations of the impact of the uncertainty. 
%{\color{red} \textit{\bfseries OPTION III for user selection:}}
%\newchanget{Preliminary simulation results available in \cite{Foad_thesis} suggest that these performance advantages
%of the proposed methods are maintained in systems that employ user selection schemes.}

The proposed \newchanget{robust power loading} techniques have been developed and evaluated in the context of a single-cell downlink with a modest
number of transmitter antennas, \revtnine{but they can} be extended in a straightforward way to the multi-cell downlink scenario \revtnine{with centralized precoder design. In principle, they can also be extended to the 
``Massive MIMO'' downlink scenario, but \revtten{for reasons of computational cost} such extensions would likely focus on the class of hybrid analog-digital beamforming
strategies for Massive MIMO; e.g., \cite{Heath_hybridBF}.}} 

\appendices

\section{Choice of $\eta_k$}
\label{app:approx}
%In the ZF beamforming case, $\operatorname{Pr}\bigl(\mathsf{SINR_k} \geq \gamma_k \bigr) \geq 1- \epsilon_k$, can be written as
%%4.16
%\begin{equation}
%\label{actual_prob_4}
%\operatorname{Pr} \Bigl(\boldsymbol{\delta}^H_k {\mathbf Q}_k\boldsymbol{\delta}_k+({\tfrac {p_k}{\gamma_k}})2\operatorname{Re}(\boldsymbol{\delta}^H_k \tilde{\mathbf {r}}_k) + v_k  \geq  0\Bigr) \geq 1-\epsilon_k.
%\end{equation}
%In this section, we have approximated this constraint by replacing the term $2\operatorname{Re}(\boldsymbol{\delta}^H_k \tilde{\mathbf {r}}_k)$ by a constant $\eta_k$. That is, we have employed the constraint
%%4.17
%\begin{equation}
%\label{appr_prob_4}
%\operatorname{Pr} \Bigl(\boldsymbol{\delta}^H_k {\mathbf Q}_k\boldsymbol{\delta}_k+({\tfrac {p_k}{\gamma_k}})\eta_k + v_k  \geq  0\Bigr) \geq 1-\epsilon_k.
%\end{equation}
%In this {\color{red}sub?}section, we seek guidelines for the choice of $\eta_k$.

One side of the tradeoff in the choice of $\eta_k$ arises from the observation that 
 %To begin, we observe that 
 the probability on the left hand side of \eqref{appr_prob_4} is a increasing function of $\eta_k$. 
 The other side arises from the observation that $\rho_k(\eta_k) = \operatorname{Pr}\bigl(2\operatorname{Re}(\boldsymbol{\delta}^H_k \tilde{\mathbf {r}}_k)\geq \eta_k \bigr)$
 %, the probability that $2\operatorname{Re}(\boldsymbol{\delta}^H_k \tilde{\mathbf {r}}_k) \geq \eta_k$ 
 is a decreasing function of $\eta_k$.
%To analyze this in more detail, in Appendix~\ref{App:lower bound} we show that if we are able to design power allocation such that \begin{math}
%\operatorname{Pr} \Bigl(\boldsymbol{\delta}^H_k {\mathbf Q}_k\boldsymbol{\delta}_k+({\tfrac {p_k}{\gamma_k}}) \eta_k+ v_k  \geq  0\Bigr) \geq 1 - \epsilon_k
%\end{math},
%then the probability that the SINR meets the specified target, $\operatorname{Pr}\bigl(\mathsf{SINR_k} \geq \gamma_k \bigr)$, can be lower bounded as 
%%4.18
%\begin{equation}
%\label{4.18}
%\operatorname{Pr}\bigl(\mathsf{SINR_k} \geq \gamma_k \bigr) = \operatorname{Pr} \Bigl(\boldsymbol{\delta}^H_k {\mathbf Q}_k\boldsymbol{\delta}_k+({\tfrac {p_k}{\gamma_k}})2\operatorname{Re}(\boldsymbol{\delta}^H_k \tilde{\mathbf {r}}_k) + v_k  \geq  0\Bigr) \geq \rho_k(\eta_k) - \epsilon_k + I^{\prime}(\eta_k),
%\end{equation}
%where $\rho_k(\eta_k) = \operatorname{Pr}\bigl(2\operatorname{Re}(\boldsymbol{\delta}^H_k \tilde{\mathbf {r}}_k)\geq \eta_k \bigr)$ and hence is a decreasing function of $\eta_k$ and $I^{\prime}(\eta_k)$ is positive increasing function of $\eta_k$.
%The lower bound in \eqref{4.18} illustrates one aspect of the trade off in selecting $\eta_k$. 
It is tempting to consider choosing $\eta_k$ so that $\rho_k(\eta_k)$ is large, as this increases the probability that designing a power loading that satisfies the approximated QoS constraint in \eqref{appr_prob_4} produces a power loading that satisfies the original QoS constraint in \eqref{eq:chance_constraint_for_original_power_load_ZF_b}. 
%Indeed, since $2\operatorname{Re}(\boldsymbol{\delta}^H_k \tilde{\mathbf {r}}_k)$ is a zero mean Gaussian random variable of variance $4\|\tilde{\mathbf{r}}_k\|^2$, it is tempting to choose $\eta_k = - 3\times 2\|\tilde{\mathbf{r}}_k\|$. 
However, doing so is inherently conservative. In particular, the approximated problem maybe infeasible when the original problem is, in fact, feasible.

Rather than fixing $\eta_k$ to a particular value, an alternative approach is to change it iteratively based on an evaluation of whether or not the current power loading satisfies the original QoS constraint
\cite{Foad_thesis}. 
Doing so provides a slight performance improvement in terms of the feasibility rate, and a significant reduction 
in the power allocated when a feasible solution is found~\cite{Foad_thesis}, but incurs the additional cost of
computing the integrals in \eqref{eq:chance_constraint_for_integral_power_load_ZF} in each cycle. 

%(This can be tested using the integral expression in \eqref{eq:chance_constraint_for_integral_power_load_ZF}, which is the specialization of the expression in \eqref{eq:chance_constraint_for_integral_power_load} to the ZF case.) If the current power loading does not satisfy the original QoS constraints, $\eta_k$ should be decreased (or $\eta^{\prime}_k$ in \eqref{eta_prim} increased), whereas if the current power loading over-satisfies the original QoS constraint $\eta_k$ can be increased (or $\eta^{\prime}_k$ in \eqref{eta_prim} decreased).  As will be seen in the experiments in Section~\ref{sec:sims}, this enables us to achieve power loadings that are close to the boundary of the feasible set and hence result in low transmission power. 

%%%%%%%%%%%%%%%%%%%%%%%%%%%%%%%%%%%%%%%%
%%%% Residue proof}
\section{Derivation of \eqref{eq:fi_hetro}}
\label{app:residue}
To prove the statement in \eqref{eq:fi_hetro} and \eqref{eq:fi_def}, we observe that
%4.8
\begin{align}
\label{eq:proof_fi_hetro_1}
%Line1
&\operatorname{Pr}\bigl(\|{\boldsymbol{\delta}}_k \|^2_{(-{{\mathbf Q}_k})}  \leq p_k/\gamma^{\prime}_k - {\sigma}^2_k\bigr) \nonumber\\
%Line2
&=  {\frac{1}{2 \pi }} \int_{-\infty}^{\infty} \frac{e^{(p_k/\gamma^{\prime}_k - {\sigma}^2_k)(\complexunit\omega+\beta)}}{\complexunit\omega+\beta} \frac{1}{\det({\mathbf I}+(\complexunit\omega+\beta)(-{\mathbf Q}_k))}d\omega\nonumber\\
%Line3
%&=  {\frac{1}{2 \pi i}} \int_{-i\infty+\beta}^{i\infty+\beta} \frac{e^{({\frac {1}{ \gamma^{\prime}_k}} p_k - {\sigma}^2_k)s}}{s} \frac{1}{\det({\mathbf I}+s(-{\mathbf Q}_k))}ds\nonumber\\
%Line4
&=  {\frac{1}{2 \pi i}} \int_{-i\infty+\beta}^{i\infty+\beta} \frac{e^{(p_k/\gamma^{\prime}_k - {\sigma}^2_k)s}}{s} \frac{1}{\prod_j (1+s{\lambda _{jk}} )}ds.
\end{align}
where $\lambda_{jk}$ is the $j^{th}$ eigenvalue of $-\mathbf{Q}_k$,
\newchanget{and, as defined after \eqref{eq:delta_probability},}
%Note that 
\begin{math}
(-{\mathbf Q}_k)  = {{\mathbf C}_k}^{1/2}\bigl( -{\frac {p_k}{\gamma_k}} {\mathbf b}_k {\mathbf b}^H_k + {{\mathbf {\bar B}}_k} {\mathbf P}{{\mathbf {\bar B}}_k}^H \bigr) {{\mathbf C}_k}^{1/2}
\end{math}, 
\newchanget{with ${\mathbf b}_k$ and ${{\mathbf {\bar B}}_k}$ containing the appropriate columns of the
matrix of %(not necessarily normalized) 
nominally zero-forcing
directions, $\mathbf{B}_{\text{ZF}}$. (Recall that $K\leq N_t$ in this case.)} 
\newchanget{When some power is allocated to each user (i.e., when each $p_k>0$), 
$(-\mathbf{Q}_k)$ is the}
%which is the 
sum of a 
 \revtsix{positive \newchanget{semi-}definite matrix of generic} 
%\newchanget{matrix that is generically positive definite
rank \newchanget{$(K-1)$} %$(r-1) = \min(N_t ,K-1)$ 
%positive definite matrix 
and a rank one negative definite matrix. 
\newchanget{The structure of these matrices and the nature of the zero-forcing directions means that for a large class of channel distributions, with high probability
$(-\mathbf{Q}_k)$ has %$r$ distinct non-zero eigenvalues. 
%In the common scenario with the number of users being at most the number of antennas (i.e., $K\leq N_t$),
% with high probability $(-\mathbf{Q}_k)$ has 
$K-1$ distinct positive eigenvalues and one negative eigenvalue.}
%Given the properties of ZF beamforming, this means that $-{\mathbf Q}_k$ typically has $r-1$ positive eigenvalues and one negative eigenvalue.  
%{\color{red}work on the writing of that statement; rather vague}
%{\color{red}so are the diagrams for eigs of Q or -Q. We need to watch out for this.}
%%%%%%%%%% Fig 4-1
%\begin{figure}
%\centering
%\includegraphics[width=0.5\figwidth]{shaded}
%\caption{General shape of the location of the poles of the integral in \eqref{eq:proof_fi_hetro_1}.}\label{fig:shaded}
%\end{figure}

\begin{figure} 
\centering
\includegraphics[width=0.5\figwidth]{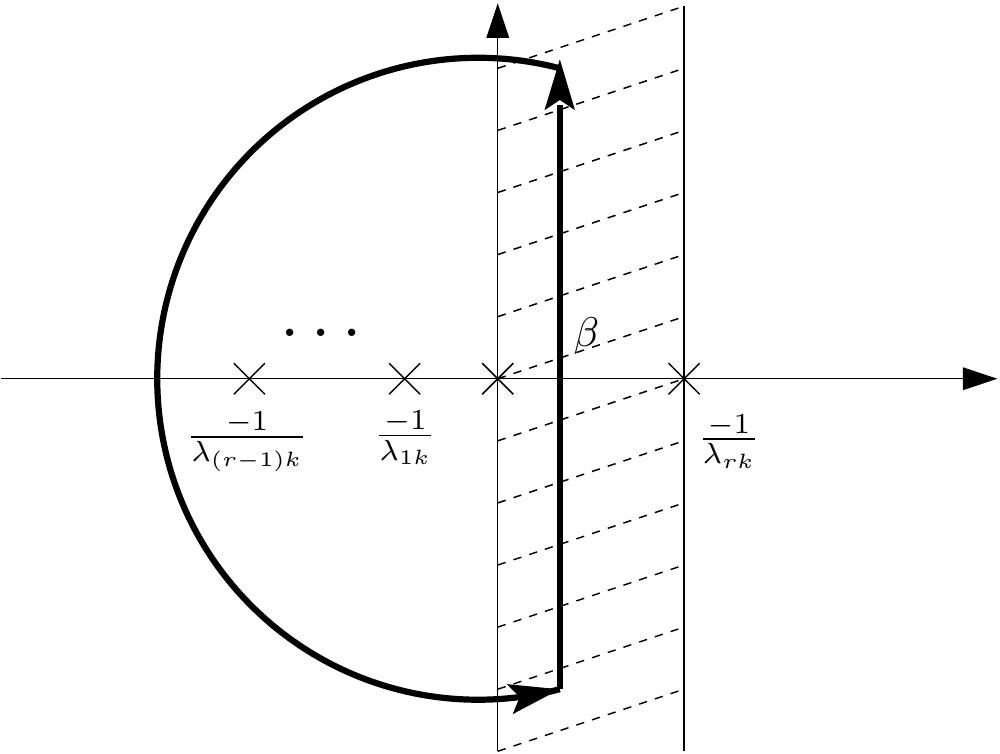} 
\hfill
\includegraphics[width=0.5\figwidth]{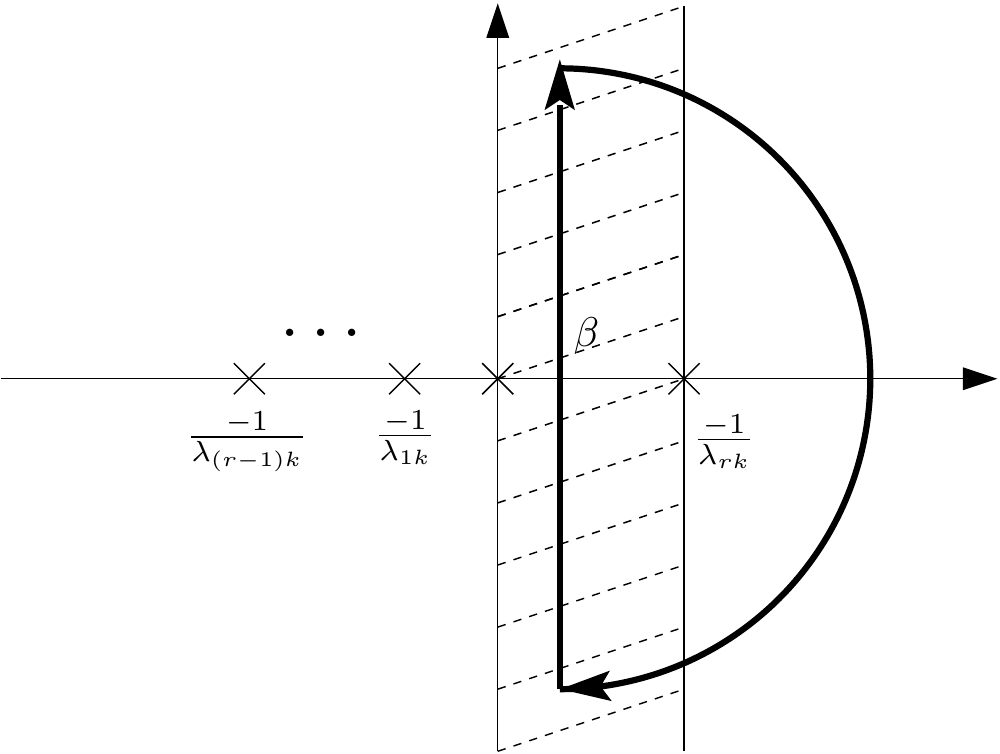}
\caption{Appropriate contour for the cases that $p_k/\gamma^{\prime}_k - {\sigma}_k^2 \geq 0$ (left) and
$p_k/\gamma^{\prime}_k - {\sigma}_k^2 < 0$ (right).}
%{\color{red}use dotted lines for the shading}} 
\label{fig:contours}
\end{figure}

\newchanget{We will denote the integrand in  \eqref{eq:proof_fi_hetro_1} by $F(s)$,
and in}
  Fig.~\ref{fig:contours}, \newchanget{we have illustrated} 
  the \newchanget{generic} shape of the location of the poles of \newchanget{$F(s)$.}
  %the argument of the integral in \eqref{eq:proof_fi_hetro_1} is illustrated.
The integral in \eqref{eq:proof_fi_hetro_1} that we are trying to evaluate is along the vertical line with real value $\beta$. According to   Lemma~\ref{lem:Hassibi}, $\beta$ can be any positive constant value that satisfies
$1+\beta {\lambda _{mk}} > 0$ for all $m$. For positive eigenvalues this constraint always holds, but for the negative eigenvalue, ${\lambda _{r k}}$, we must have $\beta < \frac{-1}{\lambda _{rk}}$. This condition and the fact that $\beta > 0$ require that $\beta$ be chosen in the 
%{\color{red}  change shading, use dotted lines for that} 
shaded region of Fig.~\ref{fig:contours}. The next step in applying residue theory is to pick the appropriate closed contour such that the integral in \eqref{eq:proof_fi_hetro_1} is equal to integral along that contour. For this purpose, a path should be added to complete the closed contour in such a way that the integral on the added path is zero. To find the appropriate path to be added, the problem in divided into two cases. In the case that $ p_k/\gamma^{\prime}_k - {\sigma}_k^2 \geq 0$, 
%{\color{red} Foad suggested mentioning $\tau_k^\prime$ here}
 the suitable contour, ${C_1}$, is depicted on the left of Fig.~\ref{fig:contours}, since the integrand is zero for the semicircular portion of the path. Therefore,  
 %when   $K=N_t$,
\newchanget{when the positive eigenvalues of $(-\mathbf{Q}_k)$ are distinct} we can use reside theory to
 %$(-\mathbf{Q}_k)$ has $(r-1)$ positive eigenvalues and they are distinct,} by using the residue theory we can 
 simplify the integral in \eqref{eq:proof_fi_hetro_1} to
%%%%%%%%%% Fig 4-2
%4.9
 \begin{align}
%& {\frac{1}{2 \pi i}} \int_{-i\infty+\beta}^{i\infty+\beta} \frac{e^{({\frac {1}{ \gamma^{\prime}_k}} p_k - {\sigma}^2_k)s}}{s} \frac{1}{\prod_j (1+s{\lambda _{jk}} )}ds\nonumber\\
%Line1
& {\frac{1}{2 \pi i}}  \oint_{C_1} F(s)ds %\nonumber\\
%line2
 =    %\underset{\text{s=0}}{\text{Res }}  
\operatorname{Res}_{s=0} F(s)
+  \sum_{\ell=1}^{\newchanget{K}-1} 
%
%\underset{\text{s= ${\frac{-1}{{\lambda}_{\ell k}}}$ }}{\text{Res }} 
\operatorname{Res}_{s=-\lambda_{\ell k}^{-1}}
F(s)\nonumber\\
%Line3
&\quad = 1 + \sum_{\ell=1}^{\newchanget{K} -1}  \lim_{s \to -\lambda_{\ell k}^{-1}} \bigl(s+\lambda_{\ell k}^{-1}\bigr)  \frac{e^{(p_k/\gamma^{\prime}_k - {\sigma}^2_k)s}}{s} \frac{1}{\prod_j (1+s{\lambda _{jk}} )}\nonumber\\
%Line4
%&= 1 +   \sum_{\ell=1}^{r -1} {-e^{({\frac {1}{ \gamma^{\prime}_k}} p_k - {\sigma}^2_k){\frac{-1}{{\lambda _{{\ell}k}}}}}} \frac{1}{\prod_{j \not= \ell} (1 - \frac{{\lambda _{jk}}}{{\lambda _{{\ell}k}}})} \nonumber\\
%Line5
&\quad = 1 + \sum_{\ell=1}^{\newchanget{K} - 1} {{f}_\ell}_k({\mathbf{P}}),
\end{align}
where ${f_\ell}_k(\cdot)$ was defined in \eqref{eq:fi_def}. 
%\newchanget{When $K<N_t$, $(-\mathbf{Q}_k)$ generically has zero eigenvalues. That case  requires a little more care, but can be handled in an analogous manner.}

On the other hand, for the case that $ p_k/\gamma^{\prime}_k - {\sigma}_k^2 <  0$,
%{\color{red}(that is, $\tau_k^\prime <0$)}, 
the suitable contour, $C_2$, is depicted on the right of Fig.~\ref{fig:contours}, since the integrand is zero for semicircular portion of that path. So in this case the residue theory should be applied just one pole. Note that this time the contour is clockwise, 
%so according to residue theory, 
so the integral along the contour is equal to the negative of the residue. In this setting the integral in \eqref{eq:proof_fi_hetro_1} can be simplified as follows
%%%%%%%%%% Fig 4-3
%\begin{figure}
%\centering
%\includegraphics[width=0.5\figwidth]{contour1}
%\caption{Appropriate contour for the cases that $p_k/\gamma^{\prime}_k - {\sigma}_k^2 < 0$.} \label{fig:contour2}
%\end{figure}
%4.10
 \begin{align}
 %& {\frac{1}{2 \pi i}} \int_{-i\infty+\beta}^{i\infty+\beta} \frac{e^{({\frac {1}{ \gamma^{\prime}_k}} p_k - {\sigma}^2_k)s}}{s} \frac{1}{\prod_j (1+s{\lambda _{jk}} )}ds\nonumber\\
%Line1
&{\frac{1}{2 \pi i}}  \oint_{C_2} F(s)ds % \nonumber\\
%Line2
 =  - %\underset{\text{s= ${\frac{-1}{{\lambda}_{rk}}}$ }}{\text{Res }} 
 \operatorname{Res}_{s=-\lambda_{r k}^{-1}} F(s)\nonumber\\
%Line3
&\quad = - \lim_{s \to -\lambda_{rk}^{-1}} \bigl(s+\lambda_{r k}^{-1}\bigr)  \frac{e^{(p_k/\gamma^{\prime}_k - {\sigma}^2_k)s}}{s} \frac{1}{\prod_j (1+s{\lambda _{jk}} )}\nonumber\\
%Line4
%&= {e^{({\frac {1}{ \gamma^{\prime}_k}} p_k - {\sigma}^2_k){\frac{-1}{{\lambda _{{r}k}}}}}} \frac{1}{\prod_{j \not= r} (1 - \frac{{\lambda _{jk}}}{{\lambda _{{r}k}}})} \nonumber\\
%Line5
&\quad = - {{f}_{r}}_k({\mathbf{P}}).
\end{align}

\begin{IEEEbiography}[{\includegraphics[width=1in,height=1.25in,clip,keepaspectratio]{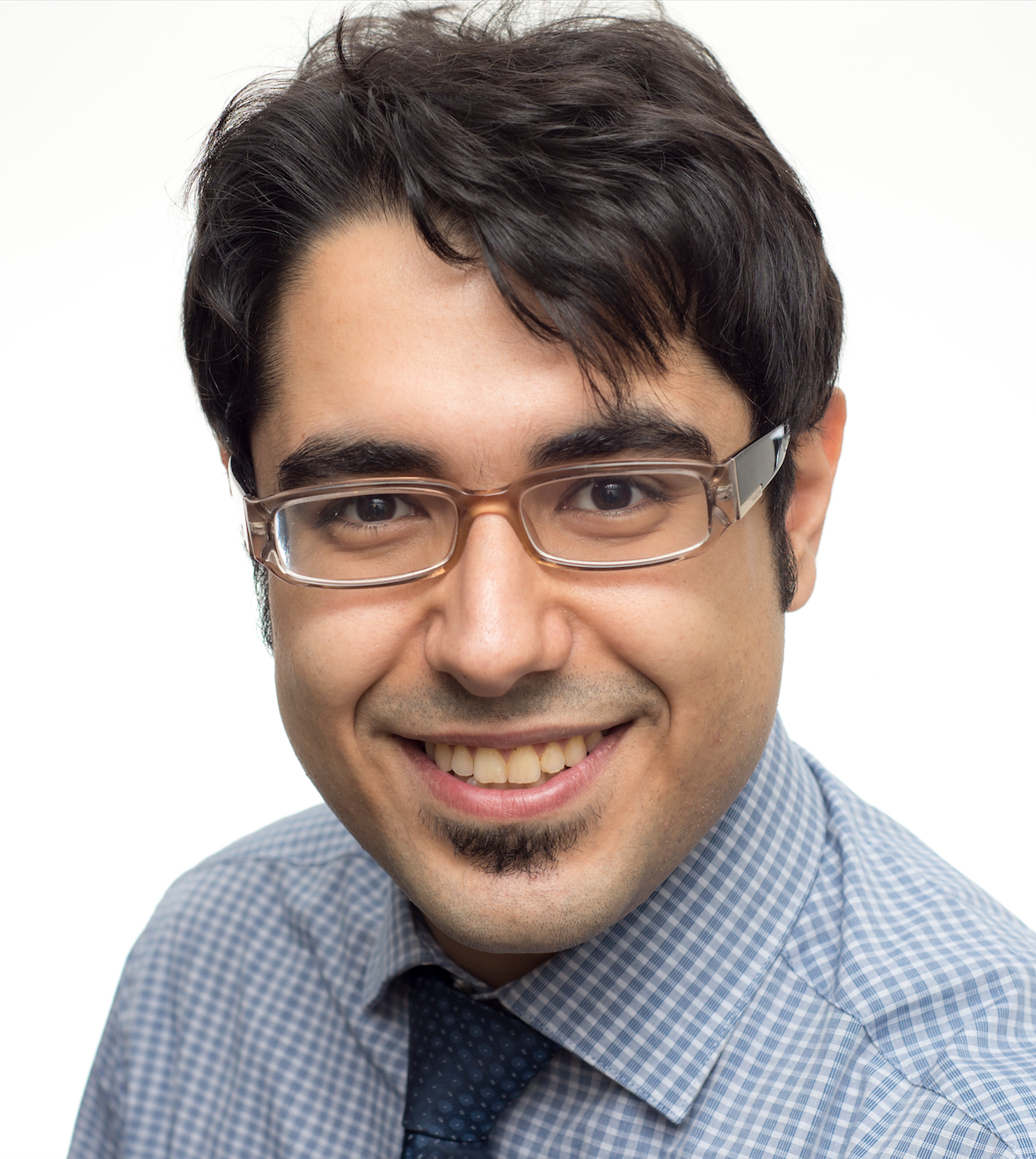}}]{Foad Sohrabi}
(S'13) received his B.A.Sc.\ degree in 2011 from the University of Tehran, Tehran, Iran, and his M.A.Sc.\ degree in 2013 from McMaster University, Hamilton, ON, Canada, both in Electrical and Computer Engineering. Since September 2013, he has been a Ph.D student at University of Toronto, Toronto, ON, Canada. Form July to December 2015, he was a research intern at Bell-Labs, Alcatel-Lucent, in Stuttgart, Germany. His main research interests include MIMO communications, optimization theory, wireless communications, and signal processing.
\end{IEEEbiography}
\begin{IEEEbiography}[{\includegraphics[width=1in,height=1.25in,clip,keepaspectratio]{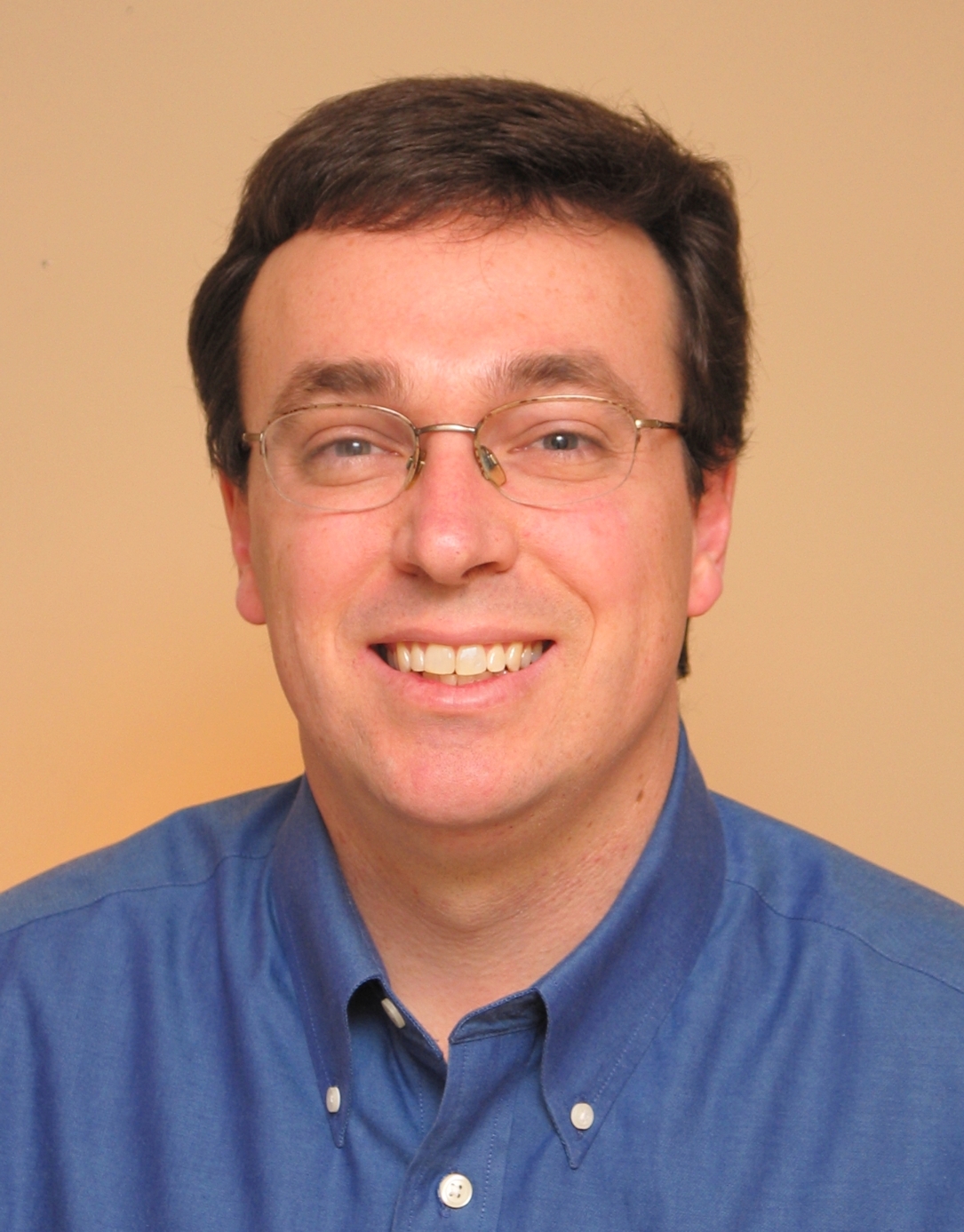}}]{Timothy~N.~Davidson} (M'96, SM'15) 
received the B.Eng. (Hons. I) degree in electronic engineering from the University of Western Australia (UWA), Perth, in 1991 and the 
D.Phil. degree in engineering science from the University of Oxford, U.K., in 1995.

He is 
a Professor in the Department of Electrical and Computer Engineering, McMaster University, Hamilton, ON, Canada, 
where he is currently serving as Chair of the Department. 
Previously, he has served as Acting Director of the School of Computational Engineering and Science for two years, and 
as
Associate Director for three years. 
His research interests lie in the general areas of communications, signal processing, and control.

Dr. Davidson received the 1991 J. A. Wood Memorial Prize from UWA, the 1991 Rhodes Scholarship for Western Australia, 
and a 2011 Best Paper Award from the IEEE Signal Processing Society. He has served as an 
Associate Editor of the \textsc{IEEE TRANSACTIONS ON SIGNAL PROCESSING}, 
the \textsc{IEEE TRANSACTIONS ON WIRELESS COMMUNICATIONS}, 
and the \textsc{IEEE TRANSACTIONS ON CIRCUITS AND SYSTEMS II}. 
He has also served as a Guest Co-Editor of issues of the 
\textsc{IEEE JOURNAL ON SELECTED AREAS IN COMMUNICATIONS}, the \textsc{IEEE JOURNAL OF SELECTED TOPICS IN SIGNAL PROCESSING}, and
the \textit{EURASIP Journal on Advances in Signal Processing}.
He was a General Co-Chair for the 2014 IEEE International Workshop on Signal Processing Advances
in Wireless Communications,  a Technical Program Co-Chair for the 2014 IEEE Global Conference
on Signal and Information Processing, and the Technical Chair for the 2015 Asilomar Conference
on Signals, Systems and Computers. 
Dr. Davidson has also served as the
Chair of the IEEE Signal Processing 
Society's 
Technical Committee on Signal Processing for Communications and Networking. He is a Registered Professional Engineer in the Province of Ontario.
\end{IEEEbiography}


\begin{thebibliography}{10}
\providecommand{\url}[1]{#1}
\csname url@samestyle\endcsname
\providecommand{\newblock}{\relax}
\providecommand{\bibinfo}[2]{#2}
\providecommand{\BIBentrySTDinterwordspacing}{\spaceskip=0pt\relax}
\providecommand{\BIBentryALTinterwordstretchfactor}{4}
\providecommand{\BIBentryALTinterwordspacing}{\spaceskip=\fontdimen2\font plus
\BIBentryALTinterwordstretchfactor\fontdimen3\font minus
  \fontdimen4\font\relax}
\providecommand{\BIBforeignlanguage}[2]{{%
\expandafter\ifx\csname l@#1\endcsname\relax
\typeout{** WARNING: IEEEtran.bst: No hyphenation pattern has been}%
\typeout{** loaded for the language `#1'. Using the pattern for}%
\typeout{** the default language instead.}%
\else
\language=\csname l@#1\endcsname
\fi
#2}}
\providecommand{\BIBdecl}{\relax}
\BIBdecl

\bibitem{Winters_Salz_Gitlin}
J.~H. Winters, J.~Salz, and R.~D. Gitlin, ``The impact of antenna diversity on
  the capacity of wireless communication systems,'' \emph{IEEE Trans.\
  Commun.}, vol.~42, no. 2/3/4, pp. 1740--1751, Feb./Mar./Apr. 1994.

\bibitem{Weingarten_2006}
H.~Weingarten, Y.~Steinberg, and S.~Shamai, ``The capacity region of the
  {G}aussian multiple-input multiple-output broadcast channel,'' \emph{IEEE
  Trans.\ Inf.\ Theory}, vol.~52, no.~9, pp. 3936--3964, Sep. 2006.

\bibitem{gesbert2007shifting}
D.~Gesbert, M.~Kountouris, R.~W. Heath, C.-B. Chae, and T.~Salzer, ``Shifting
  the {MIMO} paradigm,'' \emph{IEEE Signal Process.\ Mag.}, vol.~24, no.~5, pp.
  36--46, 2007.

\bibitem{caire2003achievable}
G.~Caire and S.~Shamai, ``On the achievable throughput of a multiantenna
  {G}aussian broadcast channel,'' \emph{IEEE Trans.\ Inf.\ Theory}, vol.~49,
  no.~7, pp. 1691--1706, 2003.

\bibitem{windpassinger2004precoding}
C.~Windpassinger, R.~F. Fischer, T.~Vencel, and J.~B. Huber, ``Precoding in
  multiantenna and multiuser communications,'' \emph{IEEE Trans.\ Wireless
  Commun.}, vol.~3, no.~4, pp. 1305--1316, 2004.

\bibitem{fung2007precoding}
C.-H. Fung, W.~Yu, and T.~J. Lim, ``Precoding for the multiantenna downlink:
  {M}ultiuser {SNR} gap and optimal user ordering,'' \emph{IEEE Trans.\
  Commun.}, vol.~55, no.~1, pp. 188--197, 2007.

\bibitem{liu2008novel}
J.~Liu and W.~A. Krzymien, ``A novel nonlinear joint transmitter-receiver
  processing algorithm for the downlink of multiuser {MIMO} systems,''
  \emph{IEEE Trans.\ Veh.\ Technol.}, vol.~57, no.~4, pp. 2189--2204, 2008.

\bibitem{hochwald2005vector}
B.~M. Hochwald, C.~B. Peel, and A.~L. Swindlehurst, ``A vector-perturbation
  technique for near-capacity multiantenna multiuser communication---{P}art
  {II}: Perturbation,'' \emph{IEEE Trans.\ Wireless Commun.}, vol.~53, no.~3,
  pp. 537--544, 2005.

\bibitem{Rashid-Farrokhi_1998_COM}
F.~Rashid-Farrokhi, L.~Tassiulas, and K.~J.~R. Liu, ``Joint optimal power
  control and beamforming in wireless networks using antenna arrays,''
  \emph{IEEE Trans. Commun.}, vol.~46, no.~10, pp. 1313--1324, Oct. 1998.

\bibitem{Beamforming_Bengtsson_2001}
M.~Bengtsson and B.~Ottersten, ``Optimal and suboptimal transmit beamforming,''
  in \emph{Handbook of Antennas in Wireless Communications}, L.~C. Godara,
  Ed.\hskip 1em plus 0.5em minus 0.4em\relax CRC Press, 2001, ch.~18.

\bibitem{spencer2004zero}
Q.~H. Spencer, A.~L. Swindlehurst, and M.~Haardt, ``Zero-forcing methods for
  downlink spatial multiplexing in multiuser {MIMO} channels,'' \emph{IEEE
  Trans.\ Signal Process.}, vol.~52, no.~2, pp. 461--471, 2004.

\bibitem{Schubert_Boche_2004}
M.~Schubert and H.~Boche, ``Solution of the multiuser downlink beamforming
  problem with individual {SINR} constraints,'' \emph{IEEE Trans. Veh. Tech.},
  vol.~53, no.~1, pp. 18--28, Jan. 2004.

\bibitem{Wiesel_2006_Fixed}
A.~Wiesel, Y.~Eldar, and S.~Shamai, ``Linear precoding via conic optimization
  for fixed {MIMO} receivers,'' \emph{IEEE Trans.\ Signal Process.}, vol.~54,
  no.~1, pp. 161--176, Jan. 2006.

\bibitem{Schubert_Boche_2007}
M.~Schubert and H.~Boche, ``A generic approach to {QoS}-based transceiver
  optimization,'' \emph{IEEE Trans. Commun.}, vol.~55, no.~8, pp. 1557--1566,
  Aug. 2007.

\bibitem{Hunger_Joham_2010}
R.~Hunger and M.~Joham, ``A complete description of the {QoS} feasibility
  region in the vector broadcast channel,'' \emph{IEEE Trans. Signal Process.},
  vol.~58, no.~7, pp. 3870--3878, Jul. 2010.

\bibitem{Huang_Palomar_separable}
Y.~Huang and D.~P. Palomar, ``Rank-constrained separable semidefinite
  programming with applications to optimal beamforming,'' \emph{IEEE Trans.
  Signal Process.}, vol.~58, no.~2, pp. 664--678, Feb. 2010.

\bibitem{Caire_Jindal_Kobayashi_Ravindran}
G.~Caire, N.~Jindal, M.~Kobayashi, and N.~Ravindran, ``Multiuser {MIMO}
  achievable rates with downlink training and channel state feedback,''
  \emph{IEEE Trans. Info. Theory}, vol.~56, no.~6, pp. 2845--2866, Jun. 2010.

\bibitem{Xu_Andrews_Jafar}
J.~Xu, J.~G. Andrews, and S.~A. Jafar, ``{MISO} broadcast channels with delayed
  finite-rate feedback: {P}redict or observe?'' \emph{{IEEE} Trans. Wireless
  Commun.}, vol.~11, no.~4, pp. 1456--1467, Apr. 2012.

\bibitem{Michael_JSTSP}
M.~{Botros Shenouda} and T.~N. Davidson, ``Convex conic formulations of robust
  downlink precoder design with quality of service constraints,'' \emph{IEEE J.
  Sel. Topics Signal Process.}, vol.~1, no.~4, pp. 714--724, Dec. 2007.

\bibitem{KKWong_robust_downlink_SDR}
G.~Zheng, K.-K. Wong, and T.-S. Ng, ``Robust linear {MIMO} in the downlink: {A}
  worst-case optimization with ellipsoidal uncertainty regions,'' \emph{EURASIP
  J. Adv. Signal Process.}, vol. 2008, no. 609018, Jul. 2008.

\bibitem{Vucic_Boche_2009}
N.~Vucic and H.~Boche, ``Robust {QoS}-constrained optimization of dowlink
  multiuser {MISO} systems,'' \emph{IEEE Trans. Signal Process.}, vol.~57,
  no.~2, pp. 714--725, Feb. 2009.

\bibitem{Michael_QoS_MSE}
M.~{Botros Shenouda} and T.~N. Davidson, ``Non-linear and linear broadcasting
  with {QoS} requirements: {T}ractable approaches for bounded channel
  uncertainties,'' \emph{IEEE Trans. Signal Process.}, vol.~57, no.~5, pp.
  1936--1947, May 2009.

\bibitem{K_Boyd_2002}
S.~Kandukuri and S.~Boyd, ``Optimal power control in interference-limited
  fading wireless channels with outage-probability specifications,'' \emph{IEEE
  Trans.\ Wireless Commun.}, vol.~1, no.~1, pp. 46--55, Jan. 2002.

\bibitem{Chalise_2007_CC_DL}
B.~Chalise, S.~Shahbazpanahi, A.~Czylwik, and A.~B. Gershman, ``Robust downlink
  beamforming based on outage probability specifications,'' \emph{IEEE Trans.\
  Wireless Commun.}, vol.~6, no.~10, pp. 3498--3503, 2007.

\bibitem{Payaro_2007}
M.~Payar{\'{o}}, A.~Pascual-Iserte, and M.~{\'{A}}. Lagunas, ``Robust power
  allocation designs for multiuser and multiantenna downlink communication
  systems through convex optimization,'' \emph{IEEE J. Sel.\ Areas Commun.},
  vol.~25, no.~7, pp. 1390--1401, Sep. 2007.

\bibitem{Michael_Asilomar08}
M.~{Botros Shenouda} and T.~N. Davidson, ``Probabilistically-constrained
  approaches to the design of the multiple antenna downlink,'' in \emph{Conf.\
  Rec.\ 42nd Ann.\ Asilomar Conf.\ Signals, Systems, Computers}, Pacific Grove,
  CA, Oct. 2008, pp. 1120--1124.

\bibitem{Vucic_Boche}
N.~Vu{\v{c}}i{\'{c}} and H.~Boche, ``A tractable method for chance-constrained
  power control in downlink multiuser {MISO} systems with channel
  uncertainty,'' \emph{IEEE Sig.\ Proc.\ Lett.}, vol.~16, no.~5, pp. 346--349,
  2009.

\bibitem{Ken_outage}
K.-Y. Wang, A.~M.-C. So, T.-H. Chang, W.-K. Ma, and C.-Y. Chi, ``Outage
  constrained robust transmit optimization for multiuser {MISO} downlinks:
  {T}ractable approximations by conic optimization,'' \emph{{IEEE} Trans.
  Signal Process.}, vol.~62, no.~21, pp. 5690--5705, Nov. 2014.

\bibitem{Komninakis_Sayed_Kalman}
C.~Komninakis, C.~Fragouli, A.~H. Sayed, and R.~D. Wesel, ``Multi-input
  multi-output fading channel tracking and equalization using {K}alman
  estimation,'' \emph{{IEEE} Trans. Signal Process.}, vol.~50, no.~5, pp.
  1065--1076, May 2002.

\bibitem{GG_Kalman_Rx}
Z.~Liu, X.~Ma, and G.~B. Giannakis, ``Space-time coding and {K}alman filtering
  for time-selective fading channels,'' \emph{{IEEE} Trans. Commun.}, vol.~50,
  no.~2, pp. 183--186, Feb. 2002.

\bibitem{Raphaeli}
D.~Raphaeli, ``Distribution of noncentral indefinite quadratic forms in complex
  normal variables,'' \emph{{IEEE} Trans. Inf. Theory}, vol.~42, no.~3, pp.
  1002--1007, May 1996.

\bibitem{Hassibi_ISIT_2009}
T.~Y. Al-Naffouri and B.~Hassibi, ``On the distribution of indefinite quadratic
  forms in {G}aussian random variables,'' in \emph{Proc.\ Int.\ Symp.\ Info.\
  Theory}, Seoul, Jun. 2009, pp. 1744--1748.

\bibitem{Yates_interference_functions}
R.~Yates, ``A framework for uplink power control in cellular radio systems,''
  \emph{{IEEE} J. Sel. Areas Commun.}, vol.~13, no.~7, pp. 1341--1347, Sep.
  1995.

\bibitem{Grundinger}
A.~Gr{\"{u}}ndinger, R.~Bethenod, M.~Joham, M.~Reimensberger, and W.~Utschick,
  ``Optimal power allocation for the chance-constrained vector broadcast
  channel and rank-one channel approximation,'' in \emph{Proc. IEEE Int.\ Wkshp
  Signal Process. Adv. Wireless Commun.}, Darmstadt, Germany, Jun. 2013, pp.
  31--35.

\bibitem{Bjornsson_etal_2014}
E.~Bj{\"{o}}rnson, M.~Bengtsson, and B.~Ottersten, ``Optimal multiuser transmit
  beamforming: {A} difficult problem with a simple solution structure,''
  \emph{IEEE Signal Process. Mag.}, vol.~31, no.~4, pp. 142--148, Jul. 2014.

\bibitem{Biguesh_Gershman_training}
M.~Biguesh and A.~B. Gershman, ``Training-based {MIMO} channel estimation: {A}
  study of estimator tradeoffs and optimal training signals,'' \emph{IEEE
  Trans.\ Signal Process.}, vol.~54, no.~3, pp. 884--893, Mar. 2006.

\bibitem{smith2004direct}
G.~S. Smith, ``A direct derivation of a single-antenna reciprocity relation for
  the time domain,'' \emph{IEEE Trans.\ Antennas Propag.}, vol.~52, no.~6, pp.
  1568--1577, 2004.

\bibitem{kaltenberger2010relative}
F.~Kaltenberger, H.~Jiang, M.~Guillaud, and R.~Knopp, ``Relative channel
  reciprocity calibration in {MIMO}/{TDD} systems,'' in \emph{Proc.\ Future
  Network and Mobile Summit, 2010}, 2010, pp. 1--10.

\bibitem{Poor:1994:ISD:174518}
H.~V. Poor, \emph{An Introduction to Signal Detection and Estimation},
  2nd~ed.\hskip 1em plus 0.5em minus 0.4em\relax Springer-Verlag, 1994.

\bibitem{Luo_SDR_tight}
E.~Song, Q.~Shi, M.~Sanjabi, R.-Y. Sun, and Z.-Q. Luo, ``Robust
  {SINR}-constrained {MISO} downlink beamforming: {W}hen is the semidefinite
  programming relaxation tight?'' \emph{EURASIP J. Wireless Commun.\
  Networking}, vol. 2012, no. 243, 2012.

\bibitem{Swindle}
C.~B. Peel, B.~M. Hochwald, and A.~L. Swindlehurst, ``A vector-perturbation
  technique for near-capacity multiantenna multiuser communication--{P}art {I}:
  Channel inversion and regularization,'' \emph{IEEE Trans.\ Commun.}, vol.~53,
  no.~1, pp. 195--202, Jan. 2005.

\bibitem{Foad_thesis}
F.~Sohrabi, ``Robust power loading for the {TDD MISO} downlink with outage
  constraints,'' Master's thesis, McMaster University, 2013, available online:
  \url{http://digitalcommons.mcmaster.ca/opendissertations/8377/}.

\bibitem{Bertsekas_NLP}
D.~P. Bertsekas, \emph{Nonlinear Programming}, 2nd~ed.\hskip 1em plus 0.5em
  minus 0.4em\relax Athena Scientific, 1999.

\bibitem{BT_E_N_Robust_Opt}
A.~{Ben-Tal}, L.~{El Ghaoui}, and A.~Nemirovski, \emph{Robust
  Optimization}.\hskip 1em plus 0.5em minus 0.4em\relax Princeton University
  Press, 2009.

\bibitem{Stridh_etal_2006}
R.~Stridh, M.~Bengtsson, and B.~Ottersten, ``System evaluation of optimal
  downlink beamforming with congestion control in wireless communication,''
  \emph{IEEE Trans.\ Wireless Commun.}, vol.~5, no.~4, pp. 743--751, Apr. 2006.

\bibitem{Davidson_Foad_ICASSP2013}
F.~Sohrabi and T.~N. Davidson, ``Coordinate update algorithms for robust power
  loading for the {MISO} downlink with outage constraints and {G}aussian
  uncertainties,'' in \emph{Proc.\ Int.\ Conf.\ Acoust., Speech, Signal
  Process.}, Vancouver, May 2013, pp. 4769--4773.

\bibitem{hassibi2003much}
B.~Hassibi and B.~M. Hochwald, ``How much training is needed in
  multiple-antenna wireless links?'' \emph{IEEE Trans.\ Inf.\ Theory}, vol.~49,
  no.~4, pp. 951--963, 2003.

\bibitem{Heath_hybridBF}
O.~{El Ayach}, S.~Rajagopal, S.~Abu-Surra, Z.~Pi, and R.~W. {Heath, Jr.},
  ``Spatially sparse precoding in millimeter wave {MIMO} systems,''
  \emph{{IEEE} Trans. Wireless Commun.}, vol.~13, no.~3, pp. 1499--1513, Mar.
  2014.

\end{thebibliography}
\end{document}